\pdfoutput=1 
\documentclass[11pt,a4paper]{article}

\usepackage{fullpage}

\usepackage{amsmath,amssymb}
\usepackage{ipa} 
\usepackage{graphicx}
\usepackage[nosort]{cite}
\usepackage{subcaption} 
\usepackage{xcolor}

\AtBeginDocument{\allowdisplaybreaks}

\usepackage[unicode,pdfborder={0 0 0},colorlinks=false,linktocpage=true]{hyperref} 

\numberwithin{equation}{section}
\allowdisplaybreaks

\makeatletter
\g@addto@macro\bfseries{\boldmath}
\makeatother

\usepackage[bbgreekl]{mathbbol}
\DeclareSymbolFontAlphabet{\mathbbm}{bbold}
\DeclareSymbolFontAlphabet{\mathbb}{AMSb}%

\newcommand{\bbone}{{\mathbbm{1}}}

\newcommand{\cD}{{\mathcal{D}}\xspace}

\newcommand{\cF}{{\mathcal{F}}\xspace}

\newcommand{\cL}{{\mathcal{L}}\xspace}
\newcommand{\cM}{{\mathcal{M}}\xspace}
\newcommand{\cN}{{\mathcal{N}}\xspace}

\newcommand{\cV}{{\mathcal{V}}\xspace}

\newcommand{\bbN}{{\mathbb{N}}\xspace}

\newcommand{\bbR}{{\mathbb{R}}\xspace}

\newcommand{\bbZ}{{\mathbb{Z}}\xspace}

\newcommand{\ul}[1]{{\underline{#1}}}
\newcommand{\dd}{{\mathrm{d}}}

\usepackage{xspace} 

\def\E#1{\ensuremath{\mathrm{E}_{#1(#1)}}\xspace}
\def\e#1{\ensuremath{\mathfrak{e}_{#1(#1)}}\xspace}

\makeatletter

\newcommand\SO{\@ifnextchar({\@SO}{\@@SO}}
\def\@SO(#1){\ensuremath{\mathrm{SO}(#1)}\xspace}
\def\@@SO#1(#2){\ensuremath{\mathrm{SO}^{#1}\kern-1pt(#2)}\xspace}

\newcommand\so{\@ifnextchar({\@so}{\@@so}}
\def\@so(#1){\ensuremath{\mathfrak{so}(#1)}\xspace}
\def\@@so#1(#2){\ensuremath{\mathfrak{so}^{#1}\kern-1pt(#2)}\xspace}

\newcommand\CSO{\@ifnextchar({\@CSO}{\@@CSO}}
\def\@CSO(#1){\ensuremath{\mathrm{CSO}(#1)}\xspace}
\def\@@CSO#1(#2){\ensuremath{\mathrm{CSO}^{#1}\kern-1pt(#2)}\xspace}

\newcommand\cso{\@ifnextchar({\@cso}{\@@cso}}
\def\@cso(#1){\ensuremath{\mathfrak{cso}(#1)}\xspace}
\def\@@cso#1(#2){\ensuremath{\mathfrak{cso}^{#1}\kern-1pt(#2)}\xspace}

\newcommand\SU{\@ifnextchar({\@SU}{\@@SU}}
\def\@SU(#1){\ensuremath{\mathrm{SU}(#1)}\xspace}
\def\@@SU#1(#2){\ensuremath{\mathrm{SU}^{#1}\kern-1pt(#2)}\xspace}

\newcommand\su{\@ifnextchar({\@su}{\@@su}}
\def\@su(#1){\ensuremath{\mathfrak{su}(#1)}\xspace}
\def\@@su#1(#2){\ensuremath{\mathfrak{su}^{#1}\kern-1pt(#2)}\xspace}

\newcommand\SL{\@ifnextchar({\@SL}{\@@SL}}
\def\@SL(#1){\ensuremath{\mathrm{SL}(#1)}\xspace}
\def\@@SL#1(#2){\ensuremath{\mathrm{SL}^{#1}\kern-1pt(#2)}\xspace}

\let\sl\relax
\newcommand\sl{\@ifnextchar({\@sl}{\@@sl}}
\def\@sl(#1){\ensuremath{\mathfrak{sl}(#1)}\xspace}
\def\@@sl#1(#2){\ensuremath{\mathfrak{sl}^{#1}\kern-1pt(#2)}\xspace}

\newcommand\GL{\@ifnextchar({\@GL}{\@@GL}}
\def\@GL(#1){\ensuremath{\mathrm{GL}(#1)}\xspace}
\def\@@GL#1(#2){\ensuremath{\mathrm{GL}^{#1}\kern-1pt(#2)}\xspace}

\newcommand\gl{\@ifnextchar({\@gl}{\@@gl}}
\def\@gl(#1){\ensuremath{\mathfrak{gl}(#1)}\xspace}
\def\@@gl#1(#2){\ensuremath{\mathfrak{gl}^{#1}\kern-1pt(#2)}\xspace}

\makeatother

\def\PSL(#1){\ensuremath{\mathrm{PSL}(#1)}\xspace}
\def\Sp(#1){\ensuremath{\mathrm{Sp}(#1)}\xspace}
\def\Spin(#1){\ensuremath{\mathrm{Spin}(#1)}\xspace}
\def\USp(#1){\ensuremath{\mathrm{USp}(#1)}\xspace}
\def\PSU(#1){\ensuremath{\mathrm{PSU}(#1)}\xspace}
\def\U(#1){\ensuremath{\mathrm{U}(#1)}\xspace}
\def\PSO(#1){\ensuremath{\mathrm{PSO}(#1)}\xspace}
\def\ISO(#1){\ensuremath{\mathrm{ISO}(#1)}\xspace}

\def\sp(#1){\ensuremath{\mathfrak{sp}(#1)}\xspace}
\def\usp(#1){\ensuremath{\mathfrak{usp}(#1)}\xspace}
\def\iso(#1){\ensuremath{\mathfrak{iso}(#1)}\xspace}

\newcommand{\mf}[1]{{\mathfrak{#1}}}

\newcommand{\hcM}{{\hat{\mathcal M}}}

\begin{document}

\titlepage
\begin{flushright}
QMUL-PH-21-43
\end{flushright}

\vspace*{1.5cm}

\begin{center}
{\bf \Large New $\cN=1$ AdS$_4$ solutions of type IIB supergravity}

\vspace*{1cm} \textsc{%
David Berman$^a$\footnote{d.s.berman@qmul.ac.uk},
Thomas Fischbacher$^b$\footnote{tfish@google.com}, and 
Gianluca Inverso$^{a,c\,}$\footnote{gianluca.inverso@pd.infn.it}%
} \\

\vspace*{0.5cm} %
$^a$ Centre for Theoretical Physics, Department of Physics and Astronomy, \\
Queen Mary University of London, 327 Mile End
Road, London E1 4NS, UK\\

\vspace*{0.5cm} $^b$ 
Google Research, Brandschenkestrasse 110, 8002 Z\"urich, Switzerland\\

\vspace*{0.5cm} $^c$ 
INFN, Sezione di Padova \\
Via Marzolo 8, 35131 Padova, Italy

\end{center}

\vspace*{2.5cm}

\begin{abstract}
We construct analytically a new family of supersymmetric AdS$_4$ solutions of IIB supergravity, with the internal space provided by a deformed $S^5\times S^1$.
The solutions preserve $\cN=1$ supersymmetry and an SO(3) subgroup of isometries of $S^5$, which is broken to U(1) along a flat direction.
They are further parametrised by a winding number and a choice of SL(2) duality twist along the circle in an elliptic conjugacy class, thus including both globally geometric and S-fold configurations.
We identify these solutions by first constructing a new family of vacua of $D=4$, $\U(4)\ltimes\bbR^{12}$ gauged maximal supergravity and use exceptional field theory to perform the uplift to ten dimensions.
We discuss the relevance of $D=5$ Wilson loops associated to preserved and broken gauge symmetries in the construction of these classes of solutions.

\end{abstract}

\vspace*{0.5cm}

\newpage
\tableofcontents
\section{Introduction}

Until recently, four-dimensional Anti de Sitter solutions of type IIB supergravity have been relatively scarce. The earliest such supersymmetric solutions were found in \cite{Lust:2009mb,Assel:2011xz}. 
Then, by applying non-abelian T-duality as a solution generating mechanism to some particular type IIA backgrounds, the authors of  \cite{Lozano:2016wrs,PandoZayas:2017ier} augmented the set of such solutions. Further examples were found using generalised geometry techniques \cite{Solard:2016uog,Passias:2017yke}.

In the last few years, further progress has been made in constructing new families of such AdS$_4$ solutions where the internal six-dimensional manifold is given by a deformed $S^5$, times a $S^1$.
The peculiar characteristic of these new solutions is that fields acquire a non-trivial \SL(2,\bbZ) monodromy around the $S^1$. 
We refer to such backgrounds as S-folds.
A first family of this kind was presented in \cite{Inverso:2016eet}, by uplifting an $\cN=4$ AdS$_4$ vacuum of a certain $D=4$ gauged maximal supergravity with gauge group $[\SO(6)\times\SO(1,1)]\ltimes\bbR^{12}$ \cite{DallAgata:2011aa,Gallerati:2014xra}.
In these solutions the monodromy around $S^1$ belongs to a hyperbolic conjugacy class of \SL(2,\bbZ), reflecting the \SO(1,1) factor in the four-dimensional gauge group. 
Their local expression was shown to match a singular limit of certain so called Janus solutions of \SO(6) gauged maximal supergravity in five dimensions \cite{DHoker:2007zhm,DHoker:2007hhe}. This then provided the link to interface configurations of $\cN=4$ super Yang--Mills and their circle compactifications \cite{Clark:2004sb,DHoker:2006qeo,Gaiotto:2008sd,Gaiotto:2008ak}.
The CFT duals of this family of solutions were identified in \cite{Assel:2018vtq}. From the CFT perspective they can indeed be thought of as the infrared limit of the circle reduction of certain interface configurations of $\cN=4$ super Yang--Mills, or alternatively as quiver Chern--Simons theories with links involving the so-called $T[U(N)]$ theory \cite{Gaiotto:2008ak}.
These CFTs were further studied and extended in \cite{Garozzo:2018kra,Garozzo:2019hbf,Garozzo:2019ejm}.

This setup has been generalised in several directions.
Solutions with fewer (or no) preserved supercharges have been constructed, again based on hyperbolic \SL(2,\bbZ) twists \cite{Guarino:2019oct,Bobev:2019jbi,Guarino:2020gfe,Bobev:2020fon}.
Numerical studies based on subsectors of $D=5$ \SO(6) gauged maximal supergravity uncovered an even richer landscape of $\mathrm{AdS}_4\times S^5\times S^1$ solutions \cite{Arav:2020obl,Arav:2021tpk}, also including examples of globally geometric (periodic) configurations.
The conformal manifolds  and Kaluza--Klein spectra of these classes of solutions have also been analysed \cite{Arav:2021gra,Bobev:2021yya,Giambrone:2021zvp,Guarino:2021kyp,Guarino:2021hrc,Cesaro:2021tna}.

In this paper we construct a new family of AdS$_4$ solutions of IIB supergravity preserving four supercharges and having internal space given by a deformed $S^5\times S^1$. 
At the maximally symmetric point, an \SO(3) subgroup of isometries of the five-sphere is also preserved, which is broken to \U(1) along a flat direction.
We find these solutions by uplifting a new family of supersymmetric vacua of a $D=4$ gauged maximal supergravity with gauge group $\U(4)\ltimes\bbR^{12} \simeq [\SO(6)\times\SO(2)]\ltimes\bbR^{12}$, which is a close cousin of the $[\SO(6)\times\SO(1,1)]\ltimes\bbR^{12}$ that has recently received much attention.
Both these gaugings were shown in \cite{Inverso:2016eet} to arise from type IIB supergravity via a generalised Scherk--Schwarz reduction on $S^5\times S^1$, where all the $S^1$ dependence is encoded in an \SL(2) element $A(\eta)$.
Whether this element is generated by a compact or non-compact generator of \SL(2) determines if the gauge group contains an \SO(2) or \SO(1,1) factor, respectively.\footnote{A parabolic $A(\eta)$ can also be considered, giving a translation factor instead of \SO(2) or \SO(1,1).}
We will display the full type IIB uplifts of the \SO(3) invariant vacuum and briefly discuss how the flat direction can be interpreted as a non-trivial fibering of $S^5$ over $S^1$, in full analogy with other solutions in the literature \cite{Giambrone:2021zvp,Guarino:2021kyp}, but also provide a convenient and complementary $D=5$ interpretation.
The resulting ten-dimensional solutions share the same local expression and are distinguished globally by a choice of periodicity of the $S^1$ coordinate $\eta$.
Depending on this choice, these backgrounds can be either globally geometric, with all fields single-valued along $S^1$, or S-fold configurations.
Contrary the other analytic solutions in the literature, the S-folds found here involve a monodromy in an elliptic conjugacy class of \SL(2,\bbZ), rather than the hyperbolic class.

The $\U(4)\ltimes\bbR^{12}$ gauging studied here was already found to admit an unstable AdS$_4$ solution in \cite{Borghese:2013dja}.
The supersymmetric solution we present here was first found numerically, following a limiting procedure starting from an \SO(3) preserving vacuum of the $\omega$ deformed \SO(8) gauged supergravity \cite{DallAgata:2012mfj}.
This is a small part of a much vaster numerical analysis of $D=4$ gauged maximal supergravity vacua that will be presented elsewhere \cite{NUMPAPER}.

The uplift of the $D=4$ solution to type IIB supergravity is carried out using the framework of exceptional field theory \cite{Hohm:2013vpa,Hohm:2013uia,Hohm:2014fxa,Abzalov:2015ega,Musaev:2015ces,Berman:2015rcc}. Exceptional field theory provides us with an organising principle of how the metric and various $p$-form fields of supergravity combine while at the same time realising various duality symmetries in a manifest way. This is particularly useful for investigating  objects like S-folds where duality is a key ingredient of the solution. For a full review of exceptional field theory and its recent applications see \cite{Berman:2020tqn}.

Given the low amount of residual (super)symmetry, the final ten-dimensional expressions are rather convoluted.
However, a few important observations can be made.
The profiles of the axio-dilaton \eqref{axiodil profile 1}--\eqref{axiodil profile 3} and its $S^1$ dependence may be important to look for the CFT duals of this family of solutions, just as the axio-dilaton profile of the $\cN=4$ S-fold solution in \cite{Inverso:2016eet} was used in \cite{Assel:2018vtq} to construct Janus configurations of $\cN=4$ SYM preserving sixteen supercharges and admitting compactification to three dimensions with a \SL(2,\bbZ) duality twist.
Furthermore, we find that the internal metric exhibits cross-terms between $S^5$ and $S^1$ that cannot be removed by globally defined diffeomorphisms, which distinguishes it from other solutions in the literature.
Our new solutions can also be phrased as solutions of $D=5$ \SO(6) gauged maximal supergravity with AdS$_4\times S^1$ topology. 
In this case, starting from the \SO(3) invariant solutions, we can extract the (constant) warp factor and also interpret the cross-terms of the type IIB internal metric in terms of a non-trivial Wilson loop along $S^1$, associated with the singlet in the decomposition of the \SO(6) gauge connection with respect to the residual \SO(3).
This interpretation immediately suggests the existence of a flat direction breaking \SO(3) to \U(1), corresponding to turning on a second Wilson loop in $D=5$, this time associated with the gauge connection of such residual \U(1) rather than a broken symmetry.
Indeed, we straightforwardly identify such one-parameter deformation with an axionic flat direction in the $D=4$ $\U(4)\ltimes\bbR^{12}$ model and find it preserves $\cN=1$ supersymmetry.
Such flat direction is entirely analogous to similar ones found for vacua of the $[\SO(6)\times\SO(1,1)]\ltimes\bbR^{12}$ gauging \cite{Guarino:2019oct,Guarino:2020gfe,Bobev:2021yya,Giambrone:2021zvp,Guarino:2021kyp,Guarino:2021hrc}.

The residual symmetries and periodicity properties of our solutions are also shared with numerical solutions found in \cite{Arav:2020obl,Arav:2021tpk} and one may wonder whether the two are in fact the same.
We argue that this is not the case, based on the five-dimensional warp factor and the comparison of the free energy for the solutions, as well as on the presence of a non-trivial Wilson loop in our case. 

The rest of this paper is organised as follows. 
In section~\ref{sec:gsugra} we review the necessary ingredients of $D=4$ gauged supergravity and describe the new $\cN=1$ AdS$_4$ vacuum of $\U(4)\ltimes\bbR^{12}$ gauged maximal supergravity.
In section~\ref{sec:UPLIFT} we summarise the framework of exceptional field theory and generalised Scherk--Schwarz reductions, and then proceed to uplift the \SO(3) invariant four-dimensional solution to type IIB supergravity. We also give a discussion of globally geometric and S-fold configurations, valid for the uplift of any solution of the $\U(4)\ltimes\bbR^{12}$ model.
In section~\ref{sec:wilson} we discuss the non-triviality of the cross-terms in the internal metric by interpreting them as a $D=5$ Wilson loop associated with a broken symmetry, and describe how Wilson loops associated to preserved symmetries give rise to an axionic flat direction. We also discuss how to make contact with previous literature on similar axionic deformations.
We conclude in section~\ref{sec:discussion}.
The appendix contains some basic information on our parameterisation of \E7 and some relevant subgroups.

\paragraph{Note added} The same \SO(3) invariant solution of $D=4$, $\U(4)\ltimes\bbR^{12}$ gauged maximal supergravity was also found independently by Fri\eth{}rik Gautason. Discussions took place with him and Nikolay Bobev about the existence of a flat direction and its interpretation. Their analysis will be published elsewhere \cite{BobevGautasonTOAPPEAR}.

\section{A new solution of \texorpdfstring{$\mathrm U(4)\ltimes\bbR^{12}$}{U(4) ⋉ R¹²} gauged maximal supergravity}
\label{sec:gsugra}
\subsection{\texorpdfstring{$D=4$}{D=4} gauged maximal supergravity}
The field content of maximal supergravity in $D=4$ dimensions is given by a metric $g_{\mu\nu}$, vector fields $A^M_\mu$ in the $\rm 56$ representation of the global \E7 symmetry of the theory, scalar fields parametrising $\E7/\SU(8)$, as well as gravitini in the $\bf 8$ and spin 1/2 fermions in the $\bf 56$ of \SU(8).
Only 28 of the vector fields carry independent degrees of freedom, while the other half are their magnetic dual.
The (lagrangian) gaugings of maximal supergravity are captured by an embedding tensor  $\Theta_M{}^\alpha$ in the $\bf912$ of \E7 \cite{deWit:2007kvg}.
The index $\alpha$ corresponds to an adjoint ($\bf133$) index of \e7, and the role of $\Theta_M{}^\alpha$ is to select a subalgebra $\mf g$ of \e7 and couple it to the vector fields so that one can define covariant derivatives
\begin{equation}
\cD_\mu = \partial_\mu  - A^M_\mu \Theta_M{}^\alpha t_\alpha\,,
\end{equation}
where $t_\alpha$ form a basis of \e7.
We absorb the gauge coupling constant into the embedding tensor.
The embedding tensor must satisfy a quadratic constraint that guarantees closure of the gauge algebra as well as gauge invariance of $\Theta_M{}^\alpha$ itself:
\begin{equation}\label{qc}
\Theta_M{}^\alpha t_{\alpha\,N}{}^P \Theta_P{}^\beta 
+ \Theta_M{}^\alpha \Theta_N{}^\gamma f_{\alpha\gamma}{}^\beta = 0\,,
\end{equation}
where $t_{\alpha\,M}{}^N$ are the generators in the $\bf56$ representation and $f_{\alpha\gamma}{}^\beta$ the \e7 structure constants.
The embedding tensor can also be written contracted with generators $t_{\alpha\,M}{}^N$ in which case it is usually denoted by $X_{MN}{}^P$:
\begin{equation}
X_{MN}{}^P=\Theta_M{}^\alpha t_{\alpha\,N}{}^P \,.
\end{equation}

All couplings of the gauged theory are determined by supersymmetry.
In this work we will mainly be concerned with the scalar potential and the supersymmetry transformations of the fermions.
Introducing the $\E7/\SU(8)$ coset representative $\cV_M{}^{\ul N}$ (and its inverse $\cV_{\ul M}{}^{N}$), where the underlined index transforms under the local \SU(8), we define the scalar field dependent $T$-tensor
\begin{equation}
T_{\ul{MN}}{}^{\ul P} = \cV_{\ul M}{}^{M} \cV_{\ul N}{}^{N} \, 
X_{MN}{}^P  \, \cV_{P}{}^{\ul P} \,,
\end{equation}
which can be decomposed into \SU(8) representations according to
\begin{equation}\label{912 branching SU8}
\mathbf{912} \ \to\ \mathbf{36}+\overline{\mathbf{36}}+\mathbf{420}+\overline{\mathbf{420}}\,.
\end{equation}
Introducing \SU(8) indices $i,j,k,\ldots$ in the $\bf8$ and $\bar{\bf8}$, the $T$-tensor decomposes into the fermion shifts $A_{1\,ij}$ and $A_{2\,i}{}^{jkl}$ and their conjugates, with
\begin{equation}
A_{1\,ij} = A_{1\,(ij)}\,,\qquad 
A_{2\,i}{}^{jkl}=A_{2\,i}{}^{[jkl]}\,,\quad
A_{2\,i}{}^{ijk} = 0\,,
\end{equation}
and complex conjugation is given by raising/lowering of all SU(8) indices.
We use the same normalisations as \cite{deWit:2007kvg}.
The fermion shifts appear in the fermion supersymmetry transformations  (the dots correspond to terms that vanish when looking at vacuum solutions) as follows
\begin{align}
\delta\psi^i_\mu &= 2\cD_\mu \epsilon^i +\ldots+ \sqrt2 A_1^{ij} \gamma_\mu \epsilon_j   \\
\delta \chi^{ijk} &= \ldots -2 \epsilon^l A_{2\,l}{}^{ijk}\,,
\end{align}
and in the scalar potential
\begin{equation}
V =\frac1{24} A_{2\,i}{}^{jkl} A_2^i{}_{jkl}  - \frac34 A_{1\,ij} A_1^{ij} \,.
\end{equation}
The stationarity condition for the scalar potential is then given by
\begin{equation}
Q^{ijkl}+\frac{1}{24} \epsilon^{ijkl\,mnpq}Q_{mnpq} = 0\,,\qquad
Q^{ijkl} = \frac34 A_{2\,m}{}^{n[ij}A_{2\,n}{}^{kl]m} - A_1^{m[i} A_{2\,m}{}^{jkl]} \,.
\end{equation}

One can also construct a \SU(8) invariant combination of the coset representatives, which is especially convenient when looking at bosonic backgrounds.
It reads
\begin{equation}\label{gsugra genmetric}
\cM_{MN} =  \cV_{M}{}^{\ul M} \cV_{N}{}^{\ul N} \Delta_{\ul{MN}} \,,
\end{equation}
where $\Delta_{\ul{MN}}$ is the \SU(8) invariant defined by the associated Cartan involution.
The scalar potential then reads
\begin{equation}
V = \frac1{672}\left(\cM^{MN}\cM^{PQ}\cM_{RS} X_{MP}{}^R X_{NQ}{}^S 
+ 7 \cM^{MN} X_{MP}{}^Q X_{NQ}{}^P\right)\,.
\end{equation}

\subsection{The \texorpdfstring{$\mathrm U(4)\ltimes\bbR^{12}$}{U(4) ⋉ R¹²} gauging}

It will be convenient to decompose \E7 with respect to its \SL(8,\bbR) subgroup.
We introduce indices $A,\,B,\,\ldots=1,\ldots,8$ in the \SL(8,\bbR) fundamental, and use the branching $\mathbf{133}\to\mathbf{63}+\mathbf{70}$ to decompose the \e7 generators into
\begin{equation}
t^A{}_B\,,\qquad t^{ABCD}\,,
\end{equation}
where the former are traceless and the latter fully antisymmetric.

Under \SL(8,\bbR), the embedding tensor representation branches as $\mathbf{912}\to\mathbf{36}+\mathbf{36}'+\mathbf{420}+\mathbf{420}'$ and a large family of consistent models is given by the first two representations, which we identify with symmetric tensors $\theta_{AB}$ and $\xi^{AB}$ and subject to the quadratic constraint $\theta_{AC}\xi^{CB} \propto \delta_A^B$ \cite{DallAgata:2011aa}.
The resulting gauge groups are contained within \SL(8,\bbR).
The \E7 defining representation then branches as $\mathbf{56}\to\mathbf{28}+\mathbf{28}'$ which we represent in double-index notation as $V_M \to (V_{[AB]}\,,V^{[AB]})$.
We can then write the non-vanishing components of the embedding tensor as follows:
\begin{equation}
\Theta_{[AB]}{}^C{}_D = 2\,\delta_{[A}^C \theta^{\vphantom{[]}}_{B]D}\,,\qquad
\Theta^{[AB]}{}^C{}_D = 2\,\delta^{[A}_D \xi_{\vphantom{[]}}^{B]C}\,.
\end{equation}

We are interested in the gauging determined by the following choice of $\theta_{AB}$ and $\xi^{AB}$:
\begin{align}\label{theta xi U4}
\theta_{AB} &= \mathrm{diag}(1,\,1,\,1,\,1,\,1,\,1,\,0 ,\,0 ) \\\nonumber
\xi^{AB}    &= \mathrm{diag}(0 ,\,0 ,\,0 ,\,0 ,\,0 ,\,0 ,\,1,\,1) \,.
\end{align}
One easily checks that the quadratic constraint \eqref{qc} is satisfied, and that the gauge algebra is $\su(4)+\mf u(1)$ plus 12 nilpotent, commuting generators transforming in the $\mathbf6_{+1} + \mathbf{6}_{-1}$ representation of $\su(4)+\mf u(1)$. 
We therefore identify a $\U(4)\ltimes\bbR^{12}$ gauge group.\footnote{This gauging was denoted $[\SO(6)\times\SO(2)]\ltimes T^{12}$ in \cite{DallAgata:2011aa}. Gravitini however transform in the $\mathbf4+\bar{\mathbf4}$ of \su(4), hence we prefer the notation above.}

\subsection{The \texorpdfstring{$\cN=1$ $(\bbZ_2)^3$}{N=1 (Z₃)²} invariant truncation}

We consistently truncate the maximal theory to an $\cN=1$ subsector based on a $(\bbZ_2)^3$ global symmetry, reflecting a certain $\mathrm G_2$ structure orbifold of a seven-torus \cite{Aldazabal:2006up,DallAgata:2005zlf}.
The discrete symmetries are embedded into the $\bf8$ of \SL(8,\bbR) as the finite transformations\footnote{\E7 truly only contains $\SL(8,\bbR)/\bbZ_2$ as a subgroup, hence the overall signs are inconsequential. We fix the ambiguity by requiring the generators to sit in the same \SL(7) subgroup, reflecting the structure group of an internal torus in a standard Kaluza--Klein reduction from eleven dimensions. A similar sign ambiguity comes from the action of these $(\bbZ_2)^3$ on the fermions, which transform in the double cover \SU(8) of the $\SU(8)/\bbZ_2$ subgroup of \E7. The choice determines which gravitino survives the truncation, but up to conventions (fixed by comparing with the superpotential) the final model is the same.}
\begin{align}
g_1 &= \mathrm{diag}(+1,\,+1,\,+1,\,-1,\,-1,\,-1,\,-1,\,+1)\,,\\\nonumber
g_2 &= \mathrm{diag}(-1,\,-1,\,+1,\,+1,\,+1,\,-1,\,-1,\,+1)\,,\\\nonumber
g_3 &= \mathrm{diag}(-1,\,+1,\,-1,\,-1,\,+1,\,-1,\,+1,\,+1)\,,
\end{align}
and are then embedded into \E7.
No vectors survive the truncation, while the invariant scalars parametrise an $\big(\SL(2)/\SO(2)\big)^7$ coset space.
We associate the seven (positive) dilatons with the \sl(8) generators
\begin{align}
\mathsf s\ \ \ \leftrightarrow\ \ & +t^1{}_1 +t^2{}_2 +t^3{}_3 -t^4{}_4 -t^5{}_5 -t^6{}_6 -t^7{}_7      \ ,  \\\nonumber
\mathsf t_1\ \ \leftrightarrow\ \ & +t^1{}_1 -t^2{}_2 -t^3{}_3 -t^4{}_4 +t^5{}_5 +t^6{}_6 -t^7{}_7      \ ,  \\\nonumber
\mathsf t_2\ \ \leftrightarrow\ \ & -t^1{}_1 +t^2{}_2 -t^3{}_3 +t^4{}_4 -t^5{}_5 +t^6{}_6 -t^7{}_7      \ ,  \\\nonumber
\mathsf t_3\ \ \leftrightarrow\ \ & -t^1{}_1 -t^2{}_2 +t^3{}_3 +t^4{}_4 -t^5{}_5 -t^6{}_6 -t^7{}_7      \ ,  \\\nonumber
\mathsf u_1\ \ \leftrightarrow\ \ & +t^1{}_1 -t^2{}_2 -t^3{}_3 +t^4{}_4 -t^5{}_5 -t^6{}_6 +t^7{}_7      \ ,  \\\nonumber
\mathsf u_2\ \ \leftrightarrow\ \ & -t^1{}_1 +t^2{}_2 -t^3{}_3 -t^4{}_4 +t^5{}_5 -t^6{}_6 +t^7{}_7      \ ,  \\\nonumber
\mathsf u_3\ \ \leftrightarrow\ \ & -t^1{}_1 -t^2{}_2 +t^3{}_3 -t^4{}_4 -t^5{}_5 +t^6{}_6 +t^7{}_7      \ ,  \\\nonumber
\end{align}
and the associated axions $\sigma$, $\tau_a$ and $\nu_a$ ($a=1,2,3$) with
\begin{equation}
t_{ABC8}\,,\qquad[ABC]\in\{123\,,156\,,246\,,345\,,147\,,257\,,367\}\,,
\end{equation}
where each triplet corresponds to one of the axions in the same order as for the dilatons above.
We then define 
\begin{equation}
S=\mathsf s+i\,\sigma\,,\qquad 
T_a = \mathsf t_a+i\,\tau_a\,,\qquad
U_a = \mathsf u_a+i\,\nu_a\,,
\end{equation}
and take all fields canonically normalised.
Each \SL(2) factor is completed by the dual of the associated axion generator, e.g. the $S$ factor is completed by $t_{4567}$ and so on.
By also embedding $(\bbZ_2)^3$ into \SU(8), one finds that only one gravitino survives the $(\bbZ_2)^3$ invariant truncation, hence the resulting theory is $\cN=1$.
The K\"ahler potential is $K=-\sum_{X}\log(X +\bar X)$ for $X=(S,\,T_a,\,U_a)$.

The embedding tensor must also be invariant under $(\bbZ_2)^3$ for the truncation to be consistent. 
When this is the case, the superpotential of the $\cN=1$ model is identified from the only invariant eigenvalue of $A_{1\,ij}$ (more precisely, this eigenvalue equals $\frac{\kappa^2}{\sqrt2}e^{\kappa^2 K/2}W$ in the normalisation conventions of \cite{Freedman:2012zz}).
The form of this superpotential for all compatible $\cN=8$ gaugings was given in \cite{Catino:2013ppa}.
For gaugings based on diagonal $\theta_{AB}$ and $\xi^{AB}$, we have
\begin{align}
-\tfrac12 W =\ & 
 \theta_{11}\, S\, T_1 U_2 U_3 + \theta_{22}\, U_1 U_3 T_1 T_3 
+\theta_{33}\, S\, U_1 U_2 T_3 + \theta_{44}\, U_2 U_3 T_2 T_3 \\\nonumber&
+\theta_{55}\, S\, U_1 T_2 U_3 + \theta_{66}\, U_1 U_2 T_1 T_2
+\theta_{77}\, S\, T_1 T_2 T_3 + \theta_{88}\, \\\nonumber&
+i\,\xi^{11}\, T_2 T_3 U_1 + i\,\xi^{22}\, S\, T_2 U_2
+i\,\xi^{33}\, T_1 T_2 U_3 + i\,\xi^{44}\, S\, T_1 U_1\\\nonumber&
+i\,\xi^{55}\, T_1 U_2 T_3 + i\,\xi^{66}\, S\, T_3 U_3
+i\,\xi^{77}\, U_1 U_2 U_3 + i\,\xi^{88}\, S\, T_1 T_2 T_3 U_1 U_2 U_3 
\,.
\end{align}
This fully determines the $\cN=1$ model.
In particular, the K\"ahler covariant derivative of the superpotential is (we set $\kappa^2=1$)
\begin{equation}
D_X W = \frac{\partial W}{\partial X} + \frac{\partial K}{\partial X} W \,,
\end{equation}
where $X$ are again the seven complex scalar fields.
The metric on the scalar manifold is $g_{X\overline Y} = \partial_X \partial_{\overline Y} K$, and the scalar potential is given by
\begin{equation}
V = e^K\left( g^{X\overline Y} D_X W \, D_{\overline Y}\overline W  - 3 |W|^2  \right)\,.
\end{equation}

\subsection{The SO(3) invariant solution}\label{sec:SO3 inv sol}

We now focus on the $\U(4)\ltimes\bbR^{12}$ gauging defined by \eqref{theta xi U4}.
We immediately notice that we can further consistently truncate the $\cN=1$ model of the previous section to an \SO(3) invariant subsector (breaking the $\bf6$ of \SU(4) to $\mathbf3+\mathbf3$) given by
\begin{equation}
T_a = T,\,\qquad U_a = U\,.
\end{equation}
Within this subsector, imposing supersymmetry by setting $D_XW = 0$ reduces to the conditions
\begin{align}
3 \,\bar S \,T -3 \,T^2 +i\,U\big(\bar S\,T^3-1\big)=0\,,&\nonumber\\
2|T|^2 -T^2 -i\,U +S\,\big[\bar T+T\,\big(i\,|T|^2U -2\big)\big]=0\,,&\\\nonumber
\big(S+T\big)\,T\,\big(2\bar U-U\big)+i\,|U|^2\big(1+S\, T^3\big)=0\,,&
\end{align}
which have the only two solutions
\begin{align}
\label{N=1 solution 1}
S&=\frac{\sqrt5+5i}{5\,3^{1/4}\sqrt2}\,,\quad
T= \frac{\sqrt5-i}{3^{1/4}\sqrt2}\,,\quad
U= \frac{\sqrt5+2i}{\sqrt3}\,,\\[1ex]
\label{N=1 solution 2}
S&=\frac{\sqrt5-5i}{3^{3/4}\sqrt2}\,,\quad
T= \frac{\sqrt5+i}{3^{3/4}\sqrt2}\,,\quad
U= \frac{\sqrt5+2i}{\sqrt3}\,.
\end{align}
Both of these solutions are indeed stationary points of the scalar potential of the full $\cN=8$ theory. 
An explicit parametrisation of the full $\E7/\SU(8)$ coset representative is described in appendix~\ref{app:e7 param}.
The value of the potential at the extrema equals
\begin{equation}
\Lambda_{\rm cosm} = -\frac{ 243 \sqrt3 }{ 25 \sqrt5 }\,.
\end{equation}
Thus, these are indeed new $\cN=1$ AdS$_4$ solutions in the full maximal supergravity.

We now summarise the mass spectrum at these vacua.
The scalar field masses are given as $(m_{\rm scal}^2 L^2)^{\rm multipl.}$ with $\Lambda_{\rm cosm}=-3/L^2$:
\begin{equation}\label{scalar masses}
\begin{aligned}
&
0^{\times28}                        \,,\ \,
-\frac{20}{9}^{\times5}             \,,\ \,
-2 ^{\times3}                       \,,\ \,
-\frac89 ^{\times3}                 \,,\ \,
\frac43^{\times5}                   \,,\ \,
8^{\times6}                         \,,\ \,
10                                  \,,\ \,
18                                  \,,\ \,
\frac{71-3\sqrt{89}}{9}^{\times5}   \,,\ \,
\frac{71+3\sqrt{89}}{9}^{\times5}   \,,\\[1ex]&
-2.223210                  \,,\ \ \,
-1.55059                  \,,\ \ \,
 3.42004                   \,,\ \ \,
 9.18240                   \,,\ \ \,
 10.2209                   \,,\ \ \,
 18.2838 
 \,.
\end{aligned}
\end{equation}
Out of the 28 massless scalars, 25 are Goldstone bosons for the broken gauge symmetries.
The six numerical values in the second line can be written exactly as the solutions $y_*$ of a polynomial equation, which comes from factorisation of the eigenvalue equation for the scalar field mass matrix and after simplification reduces to
\begin{align}
%
  6561 \,y_*^6
- 244944 \,y_*^5 
+ 2793852 \,y_*^4 
- 7955712 \,y_*^3 &\\[0.5ex]\nonumber
- 28909760 \,y_*^2 
+ 71795200 \,y_* 
+ 132736000  &\ =\ 0
\,.
\end{align}
This is the minimal polynomial encoding the associated mass eigenvalues.
The vector field masses $(m_{\rm vec}^2L^2)^{\rm multipl.}$ are
\begin{equation}\label{vector masses}
\begin{aligned}
0^{\times 3}\,,\ \ 
2^{\times4} \,,\ \ 
\frac{41-3\sqrt{41}}{9}^{\times 3} \,,\ \ 
\frac{10}{3}^{\times 5} \,,\ \ 
6^{\times 4} \,,\ \ 
\frac{41+3\sqrt{41}}{9}^{\times 3} \,,\ \ 
10^{\times 6}\,,  
\end{aligned}
\end{equation}
and we recognise the three massless vectors associated to the residual \SO(3) symmetry.
The gravitino masses $(m_{\rm 3/2}^2L^2)^{\rm multipl.}$ are
\begin{equation}\label{gravitino masses}
1\,,\quad
4^{\times4}\,,\quad
\frac{41}{9}^{\times3}\,,
\end{equation}
and we identify the first eigenvalue with the preserved supersymmetry.
Finally, the eigenvalues $(m_{\rm 1/2}^2 L^2)^{\rm multipl.}$ of the (unprojected) mass matrix for the spin 1/2 fields are 
\begin{equation}\label{spinor masses}
\begin{aligned}
&0^{\times3}\,,\ \ 
\frac{4}{9}^{\times5}\,,\ \ 
1^{\times3}\,,\ \ 
4^{\times4}\,,\ \ 
\frac{41}{9}^{\times3}\,,\ \ 
\frac{89}{9}^{\times5}\,,\ \ 
{16}^{\times5}\,,\ \ 
\frac{164}{9}^{\times3}\,,\\[1ex]&
\frac{21-\sqrt{41}}{2}^{\times6}\,,\ \ 
\frac{21+\sqrt{41}}{2}^{\times6}\,,\ \ 
\frac{23-\sqrt{129}}{6}^{\times5}\,,\ \ 
\frac{23+\sqrt{129}}{6}^{\times5}\,,\\[1ex]&
0.113103\,,\ \ 
8.30122\,,\ \ 
16.2523\,.
\end{aligned}
\end{equation}
The numerical values are the solutions $y_*$ of the equation
\mbox{$81 y_*^3-1998 y_*^2 +11153 y_* -1236 =0$}, and similar comments apply as for the numerical values in \eqref{scalar masses}.

As we anticipated, the \SO(3) invariant solutions identified here can be generalised to a one-parameter family of \U(1) invariant ones, by turning on an axionic deformation.
The relevant axion is not contained within the $\cN=1$ subsector analysed here.
We will discuss this flat direction in section~\ref{sec:wilson flat dirs} and the end result is displayed in equation~\eqref{sol with flatdir}.

\section{Uplift to type IIB supergravity}\label{sec:UPLIFT}

\subsection{IIB supergravity expressions}\label{sec:IIBsugra}

Let us briefly summarise our notation and conventions for type IIB supergravity.
The bosonic field content is given by an (Einstein frame) ten-dimensional metric, an \SL(2) doublet of two-form potentials $B_2^a$ (with $a=1,2$ in the \SL(2) fundamental), a four-form potential $C_4$ and an axio-dilaton parametrising $\SL(2)/\SO(2)$ which we can write as a complex scalar $\tau$ or in terms of a symmetric $2\times2$ matrix $m_{ab}$:
\begin{equation}\label{axio-dilaton}
\tau = C_0 + \mathrm i\, e^{-\phi}\,,\qquad 
m_{ab} = \begin{pmatrix}
e^{-\phi} + e^\phi C_0^2  &  -e^\phi C_0 \\
-e^\phi C_0               &  e^\phi 
\end{pmatrix}\,.
\end{equation}
Defining field strengths
\begin{equation}
F_3^a = \dd B_2^a\,,\qquad
F_5 = \dd C_4 -\frac12 \epsilon_{ab} B_2^a \wedge F_3^b\,,
\end{equation}
the dynamics are captured by the pseudo-action
\begin{equation}
\begin{aligned}
S_{\rm IIB} =\ &
  \int \dd^{10}\!x \sqrt{-g_{\rm IIB}} R 
+ \frac14 \int\dd m_{ab} \wedge \star \dd m^{ab} 
- \frac12 \int m_{ab} F_3^a \wedge\star F_3^b
- \frac14 \int F_5\wedge\star F_5\\&
-\frac14 \int \epsilon_{ab} C_4\wedge F_3^a \wedge F_3^b\,,
\end{aligned}
\end{equation}
combined with the self-duality constraint $F_5 = \star F_5$.

\subsection{Exceptional field theory and generalised Scherk--Schwarz reductions}\label{sec:gSS}

The maximal gauged supergravity defined by \eqref{theta xi U4} is known to descend from type IIB supergravity through a generalised Scherk--Schwarz reduction \cite{Berman:2012uy,Aldazabal:2013mya,Aldazabal:2013via,Lee:2014mla,Hohm:2014qga,Inverso:2016eet}.
This is where a Scherk--Schwarz type of reduction is carried out inside the framework of \E7 exceptional field theory \cite{Hohm:2013uia}. See \cite{Berman:2020tqn} for a review of the whole subject, and references therein. In ExFT, the spacetime is split into two parts, an external space, where we will have coordinates $x^\mu$, which is described by usual Riemannian geometry and an internal space, with coordinates $y^m$, which is described by an extended exceptional geometry.
Thus in the ExFT formulation the relevant (bosonic) fields of IIB supergravity are encoded by a four-dimensional Einstein frame metric $\hat g(x,y)_{\mu\nu}$ and a unimodular generalised metric $\hat\cM(x,y)_{MN}$, analogous to \eqref{gsugra genmetric}, parametrising a $\E7/\SU(8)$ coset.
We will not need vector fields and other $p$-forms in this paper.
All the fields depend on the external spacetime coordinates $x^\mu$ and on the internal space coordinates $y^m$.
The associated derivatives $\partial_m=\frac{\partial}{\partial y^m}$ are embedded into an exceptional internal derivative $\partial_M$ which lies in the ${\bf56}_1$ of $\E7\times\bbR^+$ according to the decomposition
\begin{equation}\label{56 IIB deco}
\mathbf{56}_1 \to    
\boxed{(\mathbf1,\mathbf{6})_{+1}}   
+ (\mathbf2,\mathbf{6}')_{-1}        
+ (\mathbf1,\mathbf{20})_{-3}        
+  (\mathbf2,\mathbf{6})_{-5}        
+ (\mathbf1,\mathbf{6}')_{-7}        
\end{equation}
under the $\SL(2)\times\GL(6)$ subgroup of $\E7\times\bbR^+$ that we identify with the global symmetry of type IIB supergravity times the \GL(6) structure group of the internal manifold.
The derivatives $\partial_m$ are mapped into the first (boxed) entry, with all other entries in $\partial_M$ vanishing. This then determines a solution (in the IIB maximal orbit) of the ExFT section constraint, which we review below.

Let us now describe some of the salient features of exceptional field theory.
The local symmetry of \E7 exceptional field theory is described by generalised diffeomorpshisms \cite{Berman:2012vc}. 
These are  generated by generalised vectors $\Lambda^M$ (in the ${\bf56}_{-1}$) acting on fields through the generalised Lie derivative as follows:
\begin{equation}
\cL_\Lambda V^M = 
\Lambda^P\,\partial_P V^M - V^P\,\partial_P \Lambda^M
-\left(12\,t_\alpha{}^{MN}t^\alpha{}_{PQ} +\frac12 \Omega^{MN}\Omega_{PQ}\right) \,\partial_N\Lambda^P\,V^Q\, .
\end{equation}
$\Omega_{MN}=\Omega^{MN}$ is the \E7 symplectic invariant (we use NW-SE conventions to raise and lower $\bf56$ indices).
Consistency of the theory, by which we mean closure of this algebra, requires us to impose a constraint on the internal derivatives $\partial_M$, which reads
\begin{equation}
t_\alpha{}^{MN} \partial_M\otimes\partial_N = 0 = \Omega^{MN}\partial_M\otimes\partial_N\,.
\end{equation}
This is sometimes called the section constraint or strong constraint \cite{Berman:2011cg,Berman:2012vc}.
This determines how the physical internal derivatives are embedded into $\partial_M$. More intuitively, the solution of the section condition describes how the physical spacetime is embedded in the extended space. This section constraint may be solved by the embedding described in \eqref{56 IIB deco} providing us with an embedding of IIB supergravity in ExFT.

We do not need to review the whole structure and dynamics of \E7 ExFT.
It suffices to state that vacuum solutions of a maximal $D=4$ gauged supergravity can be uplifted to ten- or eleven-dimensional supergravity (depending on how we solve the section constraint) if we can find a generalised frame (or twist matrix) for the internal space
\begin{equation}
E(y)_{\bar M}{}^M\ \in\ \E7\times\bbR^+\,,
\end{equation}
where the barred indices are inert under the action of the generalised Lie derivative. This twist matrix must satisfy the condition\\
\begin{equation}\label{gSS condition}
\cL_{E_{\bar M}}E_{\bar N}{}^M = -X_{\bar M \bar N}{}^{\bar P} E_{\bar P}{}^M\,,
\end{equation}
where $X_{\bar M \bar N}{}^{\bar P}$ is independent of $y$ and corresponds to the embedding tensor of the gauged supergravity.
From now on, indices of gauged supergravity will be barred (in contrast to section~2) to distinguish them from ExFT `curved' indices.
If \eqref{gSS condition} holds, vacuum solutions are uplifted by first embedding them into ExFT through the relations 
\begin{equation}\label{gss ansatz}
\begin{aligned}
\hat g(x,y)_{\mu\nu} &= \big[\det E(y)\,\big]^{1/28}\ g(x)_{\mu\nu}\,,\\
\hat \cM(x,y)_{MN}   &= 
\big[\det E(y)\,\big]^{1/28}\ E(y)_{M}{}^{\bar M} E(y)_{N}{}^{\bar N} \cM_{\bar M\bar N}\,,
\end{aligned}
\end{equation}
where $g(x)_{\mu\nu}$ is the dS, Minkowski or AdS metric of the gauged supergravity vacuum solution and $\cM_{\bar M\bar N}$ encodes the (constant) scalar vevs.
The inverse frame $E(y)_{M}{}^{\bar M}$ appears in the second line.

The expressions for the standard ten- or eleven-dimensional supergravity fields are then recovered from their embedding in ExFT.
In our case of uplifts from four dimensions to IIB supergravity, these expression were given in \cite{Inverso:2016eet} and we reproduce them here for convenience.
We use indices $m,n,\ldots=1,\ldots,6$ for the internal space and $a,b,\ldots=1,2$ for \SL(2) doublets.
Using the decomposition \eqref{56 IIB deco}, the generalised metric $\hat\cM_{MN}$ can be decomposed in several blocks:
\begin{equation}
\hcM_{MN} = \left(\begin{array}{lllll}
\hcM_{m\,n}             &\hcM_m{}^{nb}              &\hcM_{m\,n_1n_2n_3}          &\hcM_{m\,nb}           &\hcM_m{}^n                   \\
\hcM^{ma}{}_n           &\hcM^{ma\,nb}              &\hcM^{ma}{}^{n_1n_2n_3}      &\hcM^{ma}{}_{nb}       &\hcM^{ma\,n}                   \\
\hcM_{m_1m_2m_3\,n}     &\hcM_{m_1m_2m_3}{}^{nb}    &\hcM_{m_1m_2m_3\,n_1n_2n_3}  &\hcM_{m_1m_2m_3\,nb}   &\hcM_{m_1m_2m_3}{}^n           \\
\hcM_{ma\,n}            &\hcM_{ma}{}^{nb}           &\hcM_{ma\,n_1n_2n_3}         &\hcM_{ma\,nb}          &\hcM_{ma}{}^n           \\
\hcM^m{}_n              &\hcM^{m\,nb}               &\hcM^m{}_{n_1n_2n_3}         &\hcM^m{}_{nb}          &\hcM^{m\,n}                          \\
\end{array}\right)\,.
\end{equation}
Denoting by $G_{mn}$ the internal metric, $m_{ab}$ the axio-dilaton matrix, $B_{mn}{}^a$ the doublet of internal two-form potentials, and $C_{mnpq}$ the internal four-form potential, we have
\begin{equation}\label{uplift decomposition}
\begin{aligned}
\hcM^{m\,n}     &= G^{-1/2} G^{mn}\,,\\
\hcM^m{}_{nb}   &= \frac{1}{\sqrt2} G^{-1/2} G^{mp} B_{pn}{}^c \epsilon_{cb}\,,\\
\hcM_{ma\,nb}   &= \frac12 G^{-1/2} G_{mn} m_{ab}
                    + \frac12 G^{-1/2} G^{pq} B_{mp}{}^c B_{nq}{}^d \epsilon_{ca}\epsilon_{db}\,,\\
\hcM^p{}_{qmn} &= -2 \, G^{-1/2} G^{pr} \left( C_{rqmn} -\frac38 \epsilon_{ab} B_{p[q}{}^a B_{mn]}{}^b \right)\,.
\end{aligned}
\end{equation}
This suffices to identify all the non-vanishing IIB fields.

\subsection{IIB origin of \texorpdfstring{$\mathrm U(4)\ltimes\bbR^{12}$}{U(4) ⋉ R¹²} gauging}\label{sec:IIB origin}

The uplift of the gauging \eqref{theta xi U4} can be phrased in terms of a generalised frame taking values in the \SL(8,\bbR) subgroup of \E7 (times the trombone $\bbR^+$) \cite{Inverso:2016eet}.
To be more precise, we need to further separate the internal derivatives into those on $S^5$ and the one on the extra circle.
Using \SL(8,\bbR) indices, we first decompose $\partial_M$ into electric and magnetic components as 
\begin{equation}
\partial_M\to (\partial_{AB},\,\partial^{AB})\,.
\end{equation}
The internal derivatives along the $S^5$ coordinates $\varphi^I$, $I=1,\ldots,5$, are then embedded into the first half of the exceptional derivative as 
\begin{equation}
\partial_{I6}=\frac{\partial}{\partial\varphi^I}\,.
\end{equation}
The derivative along the $S^1$ direction $\eta$ is instead one of the `magnetic' components: 
\begin{equation}
\partial^{78}=\frac{\partial}{\partial\eta}\,.
\end{equation}
All other components of $\partial_M$ vanish.
This embedding of the physical derivatives shows that there is an \SL(2) subgroup of \SL(8,\bbR), acting on indices $A=7,8$, that is preserved and hence identified with the global symmetry of IIB supergravity.
The \GL(6) internal group acting on the six internal derivatives, corresponding to the decomposition \eqref{56 IIB deco}, is not entirely contained within \SL(8,\bbR) -- only a $\GL(5)$ subgroup is, as reflected by the circle coordinate being identified with a magnetic element of the $\bf56$.

We use here a different (but equivalent) presentation of the generalised frame compared to \cite{Inverso:2016eet}, which will make some statements later easier to illustrate.
The internal space associated to the uplift of a gauged supergravity is always defined by a coset space based on the (possibly centrally extended) gauge group \cite{Grana:2008yw,Inverso:2017lrz}.
In our case, we have\footnote{In fact, the equality only has to hold for the universal coverings, and discrete quotients may differ. In this case, the period of the $S^1$ need not be related to the \U(1) charges in the gauged supergravity. Also notice that we may write \SO(6) in place of \SU(4) and \SO(5) in place of \USp(4), reproducing the gauge group as it is realised on the bosons. This matters for the embedding of the coset representative in the $\bf56$ which we use in a few moments.}
\begin{equation}\label{S5S1 as gauge coset}
S^5\times S^1 = \frac{\U(4)\ltimes\bbR^{12}}{\USp(4)\ltimes\bbR^{12}} \,. 
\end{equation}
Using this fact, one introduces a coset representative $L(y)$ with transformation properties under the transitive action of the numerator in \eqref{S5S1 as gauge coset}  $L(y')= h(y)L(y)g$ with $g$ in the gauge group and $h(y)$ in the isotropy subgroup.
The Maurer--Cartan form then is decomposed as
\begin{equation}
\dd L \, L^{-1} = \mathring e_m{}^{\ul m} \,\dd y^m\, T_{\ul m} + Q_m{}^{\ul i} \,\dd y^m\, T_{\ul i}\,,
\end{equation}
with $T_{\ul m}$ and $T_{\ul i}$ a set of coset and isotropy group generators, respectively, and $\mathring e_m{}^{\ul m}$ a `reference vielbein' on the manifold.
The latter can be seen as an element of the \GL(6) structure group of the internal manifold, and as such has a natural embedding into $\E7\times\bbR^+$ following the decomposition \eqref{56 IIB deco}.
Similarly, the coset representative $L(y)$ has an embedding in \E7 because it is an element of the gauge group.
The generalised frame then takes the universal form \cite{Inverso:2017lrz}
\begin{equation}\label{universal frame}
E_{\bar M}{}^N = (L^{-1})_{\bar M}{}^{\bar N}\, (\mathring e^{-1})_{\bar N}{}^P \, C_P{}^N\,,
\end{equation}
where $C_P{}^N$ encodes part of the information on the reduction ansatz of the higher-dimensional scalars and $p$-forms (but not the internal metric) and takes values in a subgroup of $\E7\times\bbR^+$  defined by 
\begin{equation}
C_M{}^N \partial_N = \partial_M\,,
\end{equation}
with $\partial_M$ on section as described above.

In our case, all three factors in the right hand side of \eqref{universal frame} can be parametrised so that they only take values in \SL(8,\bbR) times the trombone $\bbR^+$.
This is obvious for $(L(y)^{-1})_{\bar M}{}^{\bar N}$, which is valued in the gauge group. 
In particular, we can choose it to be in \U(4) as all the $\bbR^{12}$ translations are part of the isotropy group. 
A natural choice of coset parametrisation allows us to make $\mathring e$ block-diagonal in the $S^5$ and $S^1$ directions, and hence valued in the $\GL(5)$ subgroup of \GL(6) which sits inside \SL(8,\bbR), up to a determinant factor (which will give the trombone component of the generalised frame).
Finally, $C_M{}^N$ will only encode the four-form potential on the five-sphere that generates the flux needed for its generalised parallelisation \cite{Lee:2014mla}.
These are also associated with some \SL(8,\bbR) elements as we will describe in a few moments.

We thus write our expressions in the fundamental of \SL(8,\bbR) (up to the trombone component which is easy to add back at the end).
We break it down to $\SL(6)\times\SL(2)\times\SO(1,1)$ to reflect the separation of the directions $A=7,8$ on which the IIB \SL(2) acts.
The \SL(6) here acts on $A=1,\ldots,6$.\footnote{We stress that this is not the same as the \SL(6) subgroup of the structure \GL(6,\bbR) giving rise to \eqref{56 IIB deco} -- the latter is not contained in \SL(8,\bbR). They share an \SL(5) subgroup acting on $A=I=1,\ldots,5$.}

The coset representative takes values in the $\SO(6)\times\SO(2)$ subgroup of $\SL(6)\times\SL(2)$, corresponding to the reductive part of the gauge group as realised on the bosons.
We then write in terms of $6\times6$ and $2\times2$ blocks\footnote{The barred indices $\bar A,\,\bar B$ play the same role as $\bar M,\,\bar N$ but for the fundamental of \SL(8,\bbR) -- i.e., they are inert under generalised diffeomorphisms.}
\begin{equation}\label{cosetrep parall}
L_{\bar A}{}^{\bar B} = \left(\begin{array}{ccc||c}
&&& \\
&L_{S^5}(\varphi)&& \\
&&& \\\hline\hline
&&\\[-2ex]
&&& A(\eta)\\[-2ex]
&&&
\end{array}\right)\,,\qquad
A(\eta)=g\begin{pmatrix}
\cos\eta & -\sin\eta\\ \sin\eta & \cos\eta
\end{pmatrix}\,,
\end{equation}
where $L_{S^5}(\varphi)$ parametrises $S^5=\SO(6)/\SO(5)$ in terms of a choice of coordinates $\varphi^I$ and $A(\eta)$ is just a rotation dependent on the $S^1$ coordinate $\eta$, multiplied from the left by some constant \SL(2) element $g$.
The latter does not affect the consistency of the generalised frame, which is going to reproduce the embedding tensor defined in \eqref{theta xi U4} for any choice of $g$. 
We will specify it in the next subsection when we discuss S-fold configurations.

The reference vielbein $\mathring e_I{}^{\ul I}$ on $S^5$ is obtained from $L_{S^5}$ and corresponds to a round sphere of unit radius.
The one on the circle is just 1.
The embedding in the \SL(8,\bbR) fundamental is then written as follows, this time separating blocks for $A=I=1,\ldots,5$, $A=6$, and $A=7,8$:
\begin{equation}\label{ref vielbein in SL8}
\mathring u_A{}^{\bar A} = \left(
\begin{array}{c||c}
\begin{array}{c|c}
&\\[-2.5ex]
\hspace*{1em}\mathring e^{-\frac{1}{4}} \, \mathring e_I{}^{\ul I\vphantom{\displaystyle\sum}}\hspace*{1em}\vphantom{\displaystyle\sum} & \\[-2.5ex]
&\\\hline
& \mathring e^{\frac{3}{4}\vphantom{\big[\big]}}\vphantom{\displaystyle\sum} 
\end{array}\!\!\!\!&\\\hline\hline
&\\[-2.5ex]
&\ \ \mathring e^{-\frac{1}{4}}\, \delta_a{}^b \hphantom{\ \ }  \\[-2.5ex]
&
\end{array}
\right)\ ,
\end{equation}
where we write $\mathring e$ for the determinant of the reference vielbein.
Embedding the \SL(8,\bbR) element above into the $\bf56$ of \E7, one has to add a trombone component proportional to 
the determinant $\mathring e$ in order to reproduce the correct embedding of the vielbein reproducing the decomposition \eqref{56 IIB deco}.
One then has
\begin{equation}\label{ref vielbein in 56}
\mathring e_{M}{}^{\bar M} = \mathring e^{-\frac12}\   \mathring u_{M}{}^{\bar M}\,,
\end{equation}
and similarly for the inverse $\mathring e_{\bar M}{}^{M}$.

Finally, the four-form generating the five-form flux on $S^5$ is encoded in the following matrix
\begin{equation}\label{C of frame}
C_A{}^{B} = \left(
\begin{array}{c||c}
\begin{array}{c|c}
&\\[-2.5ex]
\hspace*{1em}\delta_I{}^J\hspace*{1em}\vphantom{\displaystyle\sum} & \\[-2.5ex]
&\\\hline
C^{J}& 1\vphantom{\displaystyle\sum} 
\end{array}\!\!\!\!&\\\hline\hline
&\\[-2.5ex]
&\ \  \delta_a{}^b \hphantom{\ \ }  \\[-2.5ex]
&
\end{array}
\right)\ ,\qquad
\partial_J C^J = 4\,\mathring e\,.
\end{equation}
The generalised frame for the $\U(4)\ltimes\bbR^{12}$ gauging \eqref{theta xi U4} is obtained by combining these expressions into \eqref{universal frame}.
The embedding of \SL(8,\bbR) into \E7 is reviewed in appendix~\ref{app:e7 param}.

Incidentally, exchanging $A(\eta)$ in \eqref{cosetrep parall} with an hyperbolic or parabolic element of \SL(2), one recovers generalised parallelisations for $[\SU(4)\times\SO(1,1)]\ltimes\bbR^{12}$ and $[\SU(4)\times\bbR]\ltimes\bbR^{12}$ gaugings, respectively, while $A(\eta)=\text{constant}$ gives the \CSO(6,0,2) gauging \cite{Hohm:2014qga,Inverso:2016eet}.

\subsection{Globally geometric and S-fold configurations}\label{sec:Sfold config}

The generalised Scherk--Schwarz ansatz is generally globally consistent on the universal covering of the internal manifold, which in our case is $S^5\times \bbR$, the line being associated to the coordinate $\eta$.
The generalised frame is however clearly periodic in the $\bbR$ direction, which allows us to perform the identification $\eta\sim\eta+2\pi k$ and obtain a globally defined frame on $S^5\times S^1$, with $k\in\bbN_*$ indicating how many times the \SL(2) twist $A(\eta)$ winds around the circle.
Any solution of $\U(4)\ltimes\bbR^{12}$ gauged supergravity, and in particular the one we discuss here, can then be uplifted to a family of globally geometric solutions of IIB supergravity, parametrised by the winding number $k$.

By changing the periodicity of $\eta$ we can also consider S-fold configurations, just as discussed in \cite{Inverso:2016eet} for the uplift of $[\SU(4)\times\SO(1,1)]\ltimes\bbR^{12}$ gauged supergravity.
The idea is to patch the fields of type IIB supergravity along $S^1$ by some element of its \SL(2,\bbZ) global symmetry.
This is an example of compactifications with duality twists as discussed in \cite{Dabholkar:2002sy}.
In our case, the monodromy must be a representative of one of the elliptic conjugacy classes of \SL(2,\bbZ):
\begin{equation}\label{elliptic conj classes}
\mathfrak M_2 = \begin{pmatrix}
-1 & 0 \\ 0 & -1
\end{pmatrix}\,,\quad
\mathfrak M_3 = \begin{pmatrix}
0 & 1 \\ -1 & -1
\end{pmatrix}\,,\quad
\mathfrak M_4 = \begin{pmatrix}
0 & 1 \\ -1 & 0
\end{pmatrix}\,,\quad
\mathfrak M_6 = \begin{pmatrix}
1 & 1 \\ -1 & 0
\end{pmatrix}\,,
\end{equation}
generating $\bbZ_n$, for $n=2,3,4$ and $6$ respectively.

In practice, all expressions described in the previous sections are unchanged, except for a change in periodicity in $\eta$ and a $n$-dependent choice of the constant $g\in \SL(2)$ element appearing in the definition \eqref{cosetrep parall} of  $A(\eta)$ (which from now on we denote $g_n$), such that
\begin{equation}\label{Sfold A(eta)}
\eta \sim \eta + \frac{2\pi}{n}+2\pi k\,,\ \ k\in\bbN\,,\qquad
A(\eta)= g_n \begin{pmatrix}
\cos\eta & -\sin\eta\\ \sin\eta & \cos\eta
\end{pmatrix}\,:\ \ A(2\pi/n) = \mathfrak M_n\,.
\end{equation}
The constant \SL(2) transformation $g_n$ simply amounts to a field redefinition of the IIB fields.
It equals the identity for the globally geometric case $n=1$ as well as for $n=2$ and $4$.
For the odd cases one finds
\begin{equation}
g_3 = \begin{pmatrix}
-\frac{\sqrt3}{2}   &   -\frac12    \\
\frac12(1+\sqrt3)   &  \frac12(1-\sqrt3)
\end{pmatrix}\,,\qquad
g_6 = \begin{pmatrix}
\frac12(1-\sqrt3)   & \frac12(1+\sqrt3) \\
-\frac12            & -\frac{\sqrt3}{2}
\end{pmatrix}\,.
\end{equation}

We conclude that solutions of $\U(4)\ltimes\bbR^{12}$ gauged supergravity uplift to families of solutions of type IIB supergravity on $S^5\times S^1$, parametrised by the order $n=1,2,3,4,6$ of an elliptic \SL(2,\bbZ) twist along the circle and a winding number $k$.
The cases where $n=1$ correspond to globally geometric configurations.

\subsection{Linear realisation of SO(3) isometry}\label{sec:SO3 param}

In order to display explicit expressions for the uplift of the $\cN=1$, \SO(3) invariant solution \eqref{N=1 solution 1}, we use a parametrisation of the $S^5$ coset representative such that these residual isometries are realised linearly.

We begin by regarding $S^5$ as the product of two $S^2$ fibered over an interval, with either one of the two factors smoothly shrinking to zero size at each endpoint.
This makes an $\SO(3)\times\SO(3)$ group of isometries manifest, one for each sphere.
We are interested in linearly realising the diagonal subgroup.
We denote embedding coordinates for the two $S^2$ by $\vec P$ and $\vec Q$ respectively, with $P^2=Q^2=1$.
Introducing \SO(3) indices $\mathsf i,\,\mathsf j,\,\ldots=\mathsf 1,\mathsf 2,\mathsf 3$, we have the \SO(3) generators
\begin{equation}
(T_{\sf ij})_{\sf k}{}^{\sf l} = 2\delta_{\mathsf k [\mathsf i}\delta_{\mathsf j]}{}^{\sf l}\,,
\end{equation}
and then define the embedding coordinates as
\begin{equation}
\vec P = (0,0,1)\cdot O\,,\qquad
\vec Q = (0,0,1)\cdot e^{\omega T_{\sf 23}}\cdot O\,,\qquad O\in\SO(3)\,,
\end{equation}
where $O$ is a generic \SO(3) rotation.
The interpretation is that the sphere associated to $\vec Q$ is first rotated by $\omega$ with respect to the first sphere, then acting diagonally by arbitrary \SO(3) transformations allows to entirely cover both spheres independently.
Finally, the $S^5$ embedding coordinates are
\begin{equation}
Y= (\cos\theta \,\vec P,\, \sin\theta \, \vec Q)\,,\qquad Y^2=1\,.
\end{equation}
In the following we need a coordinate parametrisation of the \SO(3) rotation, which we pick to be
\begin{equation}
O = e^{-\gamma T_{\sf 12}} \, e^{(\alpha-\frac\pi2)T_{\sf 23}} \, e^{(\beta-\frac\pi2)T_{\sf 12}}\,,
\end{equation}
so that for instance $\vec P= (\cos\alpha\,\cos\beta,\ \cos\alpha\,\sin\beta,\ \sin\alpha)$.
The full coset representative $L_{S^5}$ is then given by the $6\times6$ matrix
\begin{equation}\label{S5 cosetrepr}
L_{S^5} = 
\begin{pmatrix}
\sin\theta\,\bbone_3 & -\cos\theta\,\bbone_3 \\ \cos\theta\,\bbone_3 & \hphantom{-}\sin\theta\,\bbone_3
\end{pmatrix}
\begin{pmatrix}
O &  \\ & \ \ e^{\omega T_{\sf 23}}\, O
\end{pmatrix}\,,
\end{equation}
with the local \SO(5) acting from the left on the first five rows.
The coordinate ranges are as follows:
\begin{equation}
\alpha\in(-\pi/2,\,\pi/2 )\,,\quad
\beta\in[0,\,2\pi)\,,\quad
\gamma\in[0,\,2\pi )\,,\quad
\omega\in(0,\,\pi)\,,\quad
\theta\in(0,\,\pi/2)\,.
\end{equation}

The rigid action of \SO(3) from the right is realised linearly (no local compensating \SO(5) transformation is necessary) and the coordinates $\theta$ and $\omega$ are \SO(3) invariant.
The associated vielbein $\mathring e_I{}^{\ul I} \dd \varphi^I$ 
reads 
\begin{align}
\mathring e^{\ul1} &= \dd\theta \,,\\
\mathring e^{\ul2} &= \cos\theta\sin\theta\ \dd\omega   \,,\\
\mathring e^{\ul3} &= \Psi^{\sf2} + \sin^2\!\theta\ \dd\omega   \,,\\
\mathring e^{\ul4} &= (\cos^2\!\theta+\cos\omega\sin^2\!\theta) \ \Psi^{\sf1} -\sin^2\!\theta\sin\omega\ \Psi^{\sf3}  \,,  \\
\mathring e^{\ul5} &= \cos\theta\sin\theta\big( (1-\cos\omega)\ \Psi^{\sf1} + \sin\omega\ \Psi^{\sf3} \big)   \,,
\end{align}
in terms of $\dd\theta$, $\dd\omega$ and the \SO(3) invariant one-forms 
\begin{equation}\label{psi123}
\begin{aligned}
\Psi^{\sf 1} &=
\sin\gamma\ \dd\alpha-\cos\alpha\cos\gamma\ \dd\beta
\,,\\
\Psi^{\sf 2} &=
\cos\gamma\ \dd\alpha +\cos\alpha\sin\gamma\ \dd\beta
\,,\\
\Psi^{\sf 3} &=
\dd\gamma-\sin\alpha\ \dd\beta
\,.
\end{aligned}
\end{equation}
The reference vielbein is extended to $S^5\times S^1$ by simply adding $e^{\ul6} = \dd\eta$ to the above expressions.
Its determinant is
\begin{equation}
\mathring e = \det(\mathring e_m{}^{\ul n}) = \det(\mathring e_I{}^{\ul I}) 
= \cos\alpha\, \sin\omega\, \cos^2\!\theta\, \sin^2\!\theta\,.
\end{equation}
From this, one can perform an integration to compute $C^I$ and complete the generalised frame.
A simple choice is 
$
C^I = -4 \,\delta^I_{\omega} \, \cos\alpha\, \cos\omega\, \cos^2\!\theta\, \sin^2\!\theta \,.
$

\subsection{Uplift of the solution}\label{sec:uplift}

Combining the expressions \eqref{gss ansatz} and \eqref{uplift decomposition} with the generalised frame constructed in \eqref{universal frame}--\eqref{C of frame}, we can in principle directly extract the type IIB supergravity solution.
However, for a generic parametrisation of the $S^5$ manifold, the resulting expressions are too complicated to handle.
There are two sources of simplification that make the computation manageable.
First, notice that plugging \eqref{universal frame} into \eqref{gss ansatz}, we have
\begin{equation}\label{asdfghjkl}
\hat \cM(x,y)_{MN}   = (\det \mathring e)\ \left(
 C^T \mathring e^{-T} L^{-T} \, \cM \, L^{-1} \mathring e^{-1} C
\right)_{MN}
\end{equation}
where $^T$ denotes transposition and $\cM$ encodes the four-dimensional solution \eqref{N=1 solution 1}.
Using the parametrisation \eqref{S5 cosetrepr} of the $S^5$ coset representative, we notice that its dependence on the coordinates $\alpha,\,\beta$ and $\gamma$ comes only through the $O\in\SO(3)$ transformation.
Since $\cM$ is \SO(3) invariant, the matrix $O$ does not contribute to the uplift.
Second, we can write the ten-dimensional fields in terms of a basis of \SO(3) invariant one-forms. 
One such basis is given by the reference vielbein $\mathring e_m{}^{\ul m}$ itself, because \SO(3) is realised linearly.
Using such a basis amounts to dressing \eqref{asdfghjkl} with $\mathring e_{\bar M}{}^{M}$ to `flatten' its indices.
This makes the computations manageable, but we then choose to use the simpler basis \eqref{psi123} of \SO(3) invariant one-forms to present our results.

Following the uplift procedure described until now, we find the IIB Einstein frame metric
\begin{equation}\label{IIB metric}
\dd s^2 = \tfrac5{18}\sqrt{\tfrac56}\,\Xi^{1/4} \, \dd s^2_{\rm AdS_4} 
+ G_{mn} \dd y^m\otimes \dd y^n \,,
\end{equation}
where $\dd s^2_{\rm AdS_4}$ is the AdS$_4$ metric of unit radius and its prefactor  determines the (inverse) warp factor in terms of the function
\begin{equation}
\Xi = 48+32\cos2\theta-4\cos^2\!\omega\sin^2\!2\theta\,.
\end{equation}
We shall present the uplift expressions in terms of the \SO(3) invariant forms \eqref{psi123} completed with 
\begin{equation}
\Psi^{\sf 4}=\dd\theta\,,\quad\Psi^{\sf 5}=\dd\omega\,,\quad\Psi^{\sf 6}=\dd\eta\,.
\end{equation}

We begin with the internal metric, which we parametrise as
\begin{equation}
G_{mn}\dd y^m \otimes \dd y^n = \Xi^{-3/4} \mathsf{G}_{\sf mn}\Psi^{\sf m} \otimes \Psi^{\sf n}
\end{equation}
with
\begin{equation}
\begin{aligned}
{\sf G}_{\sf 11} &= 
\sqrt{\tfrac56}\left(54+34\cos{2\theta}+4\cos{2\omega}\sin^2\!{\theta}-\Delta\right)
\,,\\
{\sf G}_{\sf 13} &= 
-\sqrt{\tfrac{10}{3}} \sin^2\!{\theta}\sin{2\omega} \, (5+3\cos2\theta)
\,,\\
{\sf G}_{\sf 22} &= 
\tfrac{1}{\sqrt{30}} \left( 3\,\Xi -8 (3+2\cos2\theta)^2 \right)
\,,\\
{\sf G}_{\sf 24} &= 
-\sqrt{\tfrac{2}{15}}\sin2\omega\,(3\sin2\theta+\sin4\theta)
\,,\\
{\sf G}_{\sf 25} &= 
\sqrt{\tfrac56}\left(\tfrac12\,\Xi-8\cos^2\theta\,(3+2\cos2\theta)\right)
\,,\\
{\sf G}_{\sf 26} &= 
\tfrac83\sqrt{\tfrac25}\sin\omega\, (3\sin2\theta+\sin4\theta)
\,,\\
{\sf G}_{\sf 33} &= 
4\sqrt{\tfrac{10}3} \sin^2\!\theta\sin^2\!\omega\,(3+2\cos2\theta)
\,,\\
{\sf G}_{\sf 44} &= 
\tfrac1{\sqrt{30}} \left( 3\,\Xi -8\sin^2\!\omega\,  (3+2\cos2\theta) \right)
\,,\\
{\sf G}_{\sf 45} &= 
-4\sqrt{\tfrac{10}{3}} \cos^3\!\theta \sin\theta \cos\omega \sin\omega
\,,\\
{\sf G}_{\sf 46} &= 
-\tfrac23\sqrt{\tfrac{2}{5}} \cos\omega\, ( 23 + 16 \cos2\theta +\cos4\theta )
\,,\\
{\sf G}_{\sf 55} &= 
\tfrac12\sqrt{\tfrac56} (\Xi-80\cos^4\!\theta)
\,,\\
{\sf G}_{\sf 56} &= 
\tfrac{16}{3}\sqrt{10} \cos^3\!\theta \sin\theta \sin\omega
\,,\\
{\sf G}_{\sf 66} &= 
\tfrac{1}{3\sqrt{30}}(3\,\Xi -32 \sin^2\!2\theta \sin^2\omega)
\,.
\end{aligned}
\end{equation}
\smallskip

Similarly, for the two-form potentials we write
\begin{equation}
B^a = 
3^{-1/4}\Xi^{-1} \mathsf B_{\sf mn}{}^b \big(A(\eta)^{-1}\big)_b{}^a\ \ \frac12\Psi^{\sf m} \wedge \Psi^{\sf n}
\end{equation}
with
\begin{align}&
\begin{aligned}
{\sf B}_{\sf 12}{}^1 &= 
-\tfrac{1}{\sqrt2} \left( 
13 \cos3\theta -\cos5\theta +4\cos\theta (17+\cos2\omega \sin^2\!\theta (2+\cos2\theta))  
\right)
\,,\\
{\sf B}_{\sf 14}{}^1 &= 
-{\sqrt2} \sin2\omega \, ( 7 \sin\theta +3\sin3\theta ) 
\,,\\
{\sf B}_{\sf 15}{}^1 &= 
-4{\sqrt2} \sin^2\!\theta \, \left( 5\cos\theta+\cos3\theta+4\cos2\omega\,\cos^3\!\theta \right) 
\,,\\
{\sf B}_{\sf 16}{}^1 &= 
4\sqrt{\tfrac{2}{3}} \cos^2\!\theta\,\left(
2\sin3\omega\sin^3\!\theta -\sin\omega\sin\theta(7+\cos2\theta)
\right)
\,,\\
{\sf B}_{\sf 23}{}^1 &= 
-4\sqrt2 \sin2\omega\sin^2\!\theta(4\cos\theta+\cos3\theta)
\,,\\
{\sf B}_{\sf 34}{}^1 &= 
8\sqrt2 \sin^2\!\omega\,(2\sin\theta+\sin3\theta)
\,,\\
{\sf B}_{\sf 35}{}^1 &= 
\tfrac5{\sqrt2} \sin2\omega\sin^32\theta \,(\sin\theta)^{-1}
\,,\\
{\sf B}_{\sf 36}{}^1 &= 
-4\sqrt{\tfrac23}\sin\omega\sin2\omega\sin\theta\sin^2\!2\theta
\,,
\end{aligned}\\&
\begin{aligned}
\\[1ex]
{\sf B}_{\sf 12}{}^2 &= 
\tfrac1{2\sqrt6}\left(
-48\cos^2\!\theta \cos3\omega \sin^3\!\theta 
+5\cos\omega\,( 46\sin\theta+11\sin3\theta-3\sin5\theta )
\right)
\,,\\
{\sf B}_{\sf 14}{}^2 &= 
2\sqrt{\tfrac23} \sin\omega\, (4\cos\theta+\cos3\theta) (3+\cos2\theta-2\cos2\omega\sin^2\!\theta)
\,,\\
{\sf B}_{\sf 15}{}^2 &= 
-\tfrac5{\sqrt6} \cos\omega\sin\theta (2\cos2\omega\sin^2\!2\theta + 5\cos4\theta-21)
\,,\\
{\sf B}_{\sf 16}{}^2 &= 
\tfrac43\sqrt2 \sin2\omega \sin^2\!\theta\,(13\cos\theta+3\cos3\theta)
\,,\\
{\sf B}_{\sf 23}{}^2 &= 
\tfrac1{\sqrt6} \sin\omega \sin\theta \left(
106-26\cos4\theta+3\cos(4\theta-2\omega) -6\cos2\omega +3\cos(4\theta+2\omega)
\right)
\,,\\
{\sf B}_{\sf 34}{}^2 &= 
8\sqrt{\tfrac23} \cos\omega\sin^2\!\omega\sin^2\!\theta\,(4\cos\theta+\cos3\theta)
\,,\\
{\sf B}_{\sf 35}{}^2 &= 
-5\sqrt{\tfrac23}\sin^3\!\theta\left(
\sin\omega\,(23+15\cos2\theta) -2\sin3\omega\cos^2\!\theta
\right)
\,,\\
{\sf B}_{\sf 36}{}^2 &= 
-\tfrac{32}{3}\sqrt2 \sin^2\!\omega\sin^2\!\theta (4\cos\theta+3\cos3\theta)
\,.
\end{aligned}
\end{align}

We present then the internal part of the $F_5$ flux, with the external part following by self-duality.
In the usual $\Psi^{\sf m}$ basis, we write
\begin{equation}
F_{5} = 3\,\Xi \, \mathsf F_{\sf mnpqr} \, (1+\star)\frac{1}{5!}
\Psi^{\sf m}\wedge\Psi^{\sf n}\wedge\Psi^{\sf p}\wedge\Psi^{\sf q}\wedge\Psi^{\sf r} \,,
\end{equation}
with
\begin{equation}
\begin{aligned}
\mathsf F_{\sf 12345} &=
-\left((27+8\cos2\theta)\sin^2\!2\theta \sin\omega \right) 
\,,\\
\mathsf F_{\sf 12346} &=
-2\sqrt3\,(4\sin2\theta+3\sin4\theta)\sin^2\!\omega
\,,\\
\mathsf F_{\sf 12356} &=
-\sqrt3 (3+2\cos2\theta) \sin^2\!2\theta \sin2\omega
\,,\\
\mathsf F_{\sf 13456} &=
-\tfrac{8}{\sqrt3}\cos\theta(1+4\cos2\theta)\sin^3\!\theta \sin^2\omega
\,.
\end{aligned}
\end{equation}

Finally, the axio-dilaton is encoded in the matrix $m_{ab}$ \eqref{axio-dilaton} which takes the form
\begin{equation}\label{axiodil matrix final expr}
m_{ab}=\Xi^{-1/2}A(\eta)_a{}^c A(\eta)_b{}^d \,{\sf M}_{cd}\,,\qquad
{\sf M}_{ab} = \begin{pmatrix}
\tfrac1{\sqrt3}(12+8\cos2\theta) &  -2\cos\omega\sin2\theta \\
-2\cos\omega\sin2\theta          &  4\sqrt3
\end{pmatrix} \,.
\end{equation}
In terms of the standard axion $C_0$ and dilaton $e^\phi$, one finds for $\bbZ_n$ identifications along $S^1$ with $n=1,2$ and 4 (such that $g_n=\bbone$ in \eqref{Sfold A(eta)})
\begin{equation}\label{axiodil profile 1}
\begin{aligned}
e^\phi &= \tfrac{1}{3\sqrt\Xi} \left(
4\sqrt3\,(3+2\cos2\theta\sin^2\!\eta)-6\cos\omega\sin2\eta\sin2\theta
\right)\,,\\
e^\phi\,C_0 &= -\tfrac{2}{3\sqrt\Xi} \left(
2\sqrt3\,\sin2\eta\cos2\theta-3\cos2\eta\cos\omega\sin2\theta
\right)\,,
\qquad\quad\text{for $n=1,2,4$.}
\end{aligned}
\end{equation}
For the $\bbZ_3$ and $\bbZ_6$ S-folds we must take into account the constant $g_n$ transformation in \eqref{Sfold A(eta)}.
We then find $C_0 = -e^{-\phi}/\sqrt3$ and
\begin{equation}\label{axiodil profile 2}
\begin{aligned}
e^\phi = 
\tfrac1{\sqrt{3\,\Xi\,}} \Big( 
  8(3+\cos2\theta)
  +2\sqrt3 \,\cos2\eta\, (2\cos2\theta + \cos\omega\sin2\theta)\ &\\
  -\sin2\eta\, (4\cos2\theta - 6\cos\omega\sin2\theta)&
 \Big)
\,,
\qquad\text{for $n=3$,}
\end{aligned}
\end{equation}
and
\begin{equation}\label{axiodil profile 3}
\begin{aligned}
e^\phi = 
\tfrac1{\sqrt{3\,\Xi\,}} \Big( 
  4(3+\cos2\theta)
  +\sqrt3\,\sin2\eta\, (2\cos2\theta - \cos\omega\sin2\theta)\ &\\
  +\cos2\eta\, (2\cos2\theta + 3\cos\omega\sin2\theta)&
 \Big)
\,,
\qquad\text{for $n=6$.}
\end{aligned}
\end{equation}
These expressions stay finite on the whole coordinate patch and, by continuity, on the whole internal manifold.
The range of values of the dilaton spans between the perturbative and non-perturbative regimes. 
Following the same line of reasoning given in \cite{Assel:2018vtq}, we may still trust the supergravity solution as long as we are in a regime of slowly varying fields, since at the two-derivative level the IIB supergravity equations of motion are fixed by supersymmetry.

While most of the explicit expressions of the full ten-dimensional solution are rather uninformative, the profile of the axio-dilaton may be of relevance in identifying the profiles of the complex coupling of $\cN=4$ SYM on a circle, which upon flowing to the IR give the CFT dual of our solutions.
We also see that the internal metric contains cross-terms between $S^5$ and $S^1$.
If we reduce our ten-dimensional solution to a Janus-like solution of $D=5$ \SO(6) gauged maximal supergravity (or equivalently uplift from four to five dimensions), with topology $\mathrm{AdS}_4\times S^1$, these cross terms appear as a constant, non-vanishing five-dimensional vector field along the $\dd\eta$ direction and associated with the generator of \SO(6) that is invariant under the residual \SO(3) gauge symmetry.
Such expectation value can be locally removed by an $\eta$ dependent gauge transformation, but a global obstruction may be present.
We analyse this further in section~\ref{sec:wilson}.

We have explicitly verified that the expressions displayed in this section satisfy the equations of motion of type IIB supergravity, including the Einstein equations.

\subsection{Supersymmetry and S-folding}

In a generalised Scherk--Schwarz reduction fermions behave like scalar densities, namely their internal space dependence is entirely encoded in an overall power of the determinant of the generalised frame.
In our case this reduces to some power of $\mathring e$, which is $\eta$ independent.
This is important in order to ensure that the residual supersymmetry of the $D=4$ solution is preserved in the type IIB solution even after compactifying $\eta$, especially when we employ a duality twist as described in section~\ref{sec:Sfold config}.
From the point of view of IIB supergravity we may expect that the axio-dilaton undergoing an elliptic \SL(2,\bbZ) twist \eqref{Sfold A(eta)} when going around $S^1$ should induce a local compensating \U(1) transformation on the fermions, hence requiring the latter, including Killing spinors, to depend non-trivially on $\eta$.
Naively, this appears in contrast with the generalised Scherk--Schwarz ansatz.
However, we are not required to parameterise the \SL(2)/\SO(2) sigma model of IIB supergravity in terms of the standard axio-dilaton. 
We may instead choose a coset representative $\ell_a{}^b$ that does not require compensating local \SO(2) transformations to accommodate non-trivial monodromies along $S^1$.
The obvious choice is the one dictated, indeed, by the Scherk--Schwarz ansatz.
Writing the coordinate dependence explicitly, we define
\begin{equation}\label{global sigma model param}
m(\varphi,\eta)_{ab} = \ell(\varphi,\eta)_a{}^c \ell(\varphi,\eta)_b{}^c \,,\quad 
\ell(\varphi,\eta)_a{}^b = A(\eta)_a{}^c\, {\sf v}(\varphi)_c{}^b
\end{equation}
where $\varphi^I$ denote the $S^5$ coordinates and ${\sf v}(\varphi)_c{}^b$ is a coset representative for the matrix $\mathsf M(\varphi)_{ab}$ appearing in \eqref{axiodil matrix final expr}.
{Of course, one may prefer to change local \SO(2) gauge to move back to a standard parametrisation of the axio-dilaton sigma model. In this case the necessary local \U(1) transformation (in the double cover) must be applied to the expression of the Killing spinors, too, thus inducing some $\eta$ dependence as expected.}
This discussion confirms that it is globally consistent to uplift the $D=4$ gauged supergravity Killing spinors preserved by our (or any other) solution, even in presence of S-folding.

\section{\texorpdfstring{$D=5$}{D=5} Wilson loops}\label{sec:wilson}

\subsection{One-form contribution in five dimensions}\label{sec:broken sym wilson}

We already observed that the ten-dimensional metric includes cross-terms between $S^5$ and $S^1$ and that the interpretation in $D=5$ \SO(6) gauged maximal supergravity is that one of the vectors in the gauge connection acquires a constant vev, with leg along $S^1$.
Let us now be more precise.
Using the isomorphism
\begin{equation}
\frac{\E7}{\SU(8)} \simeq \frac{(\E6\times\SO(1,1))\ltimes\bbR^{27}}{\USp(8)}\,,
\end{equation}
we identify with the last factor the 27 scalars of $D=4$ maximal supergravity arising from KK reduction of five-dimensional vector fields.
In ExFT language, fixing the exceptional derivative to be
\begin{equation}
\partial_{AB} = 0,\,\qquad 
\partial^{AB} = \delta^{AB}_{\,7\,8} \frac{\partial}{\partial\eta}\,,
\end{equation}
so that we embed $D=5$ maximal supergravity into \E7 ExFT, the $\bbR^{27}$ generators annihiliate the internal derivative, while their transposes do not as they are associated to hidden symmetries of the KK reduced $D=4$ theory.\footnote{Transposition is more formally defined as the Cartan involution singling out the local \SU(8), but in our parametrisation it really is just matrix transposition.}
Looking at the 14 scalar truncation described in section~\ref{sec:gsugra} and in the appendix, we find that the $\SL(2)^4$ group associated to $S$ and $T_a$ stabilises the choice of fifth coordinate and is therefore a subgroup of \E6.
Regarding the axions of the $U_a$ factors, instead, only $(t^{(\nu_a)})^T$ annihiliate the fifth coordinate and are therefore associated with three vectors in $D=5$.
Up to a change of local $\SO(2)^3$ gauge we then reparametrise the coset representative \eqref{N=1 coset param} as follows
\begin{equation}\label{N=1 coset param TILDE}
\cV_M{}^{\underline N} \to 
e^{-\sigma \, t^{(\sigma)}} \,{\sf s}^{\frac12 t^{(\mathsf s)}}  \ 
\prod_{a=1,2,3} e^{-\tau_a \, t^{(\tau_a)}} \,({\sf t}_a)^{\frac12 t^{(\mathsf t_a)}}\ 
\prod_{a=1,2,3}e^{-\tilde\nu_a \, \big(t^{(\nu_a)}\big)^T} \, (\tilde{\sf u}_a)^{\frac12 t^{(\mathsf u_a)}} \ ,
\end{equation} 
with 
\begin{equation}
\tilde{\sf u}_a = \frac{{\sf u}_a^2+\nu_a^2}{\sf u_a}\,,\qquad
\tilde\nu_a     = \frac{\nu_a}{{\sf u}_a^2+\nu_a^2}\,,
\end{equation}
where in particular the axions $\tilde\nu_a$ are canonically normalised.
It is then easy to convince oneself from the definitions in \eqref{sigmas of axions} that these three axions uplift to the $\dd\eta$ components of the five-dimensional vector fields giving the gauge connection of a special choice of Cartan subalgebra of \su(4), such that their diagonal combination is the singlet in the decomposition 
\begin{equation}\label{so6 branch}
\su(4) \to \so(3) + \mathbf3+ \mathbf3+ \mathbf5+ \boxed{ \mathbf1 }
\end{equation}
under the residual gauge symmetry of our solution.
From \eqref{N=1 solution 1}, uplifting our four dimensional solution to $D=5$ gauged maximal supergravity, we find that these three canonically normalised vectors take the value
\begin{equation}\label{one form vev}
A^a = \left(\frac23\right)^{3/2}\dd\eta\,,\quad a=1,2,3\,.
\end{equation}

Suppose we now construct the $D=5$ embedding tensor of $D=5$ \SO(6) gauged maximal supergravity.
We can fix its overall normalisation such that, upon KK compactification along the $\dd\eta$ circle with $A(\eta)$ duality twist, the resulting $D=4$ model matches \eqref{theta xi U4}.
We can then check whether the vev \eqref{one form vev} can be removed by a globally defined gauge transformation associated with the corresponding $\U(1)\subset\SO(6)$:
\begin{equation}
A^a \to A^a + \dd\lambda(\eta)\,,\quad a=1,2,3\,,\quad \lambda(\eta)\propto\eta+\text{constant}\,,
\end{equation}
by making sure that its action on the matter fields is single-valued on $S^1$.
We find that matter fields have integer charges under this \U(1) given our choices of normalisation, hence removing \eqref{one form vev} would require a multi-valued gauge transformation for any of the allowed choices of periodicity of $\eta$.
Because this \U(1) is a broken symmetry in our solution, using it to remove the vev \eqref{one form vev} would make some $\E6/\USp(8)$ scalar fields multi-valued along $S^1$.
Therefore \eqref{one form vev} is globally nontrivial.

We may rephrase this result by stating that the uplift of our solution to $D=5$ \SO(6) gauged maximal supergravity involves a nontrivial Wilson loop along $S^1$ associated with the \SO(3) singlet in \eqref{so6 branch}.
Using appropriate normalisations, this Wilson loop is invariant under gauge transformations globally defined on $S^1$ and therefore the fact that it differs from 1 indicates a global obstruction to gauging away the constant value of the associated vector field.

Of course, if we decompactify $S^1$ there is no obstruction to removing the vector vev.
In this case, the ten-dimensional interpretation of such procedure is to perform an $\eta$-dependent diffeomorphism on $S^5$ to remove the cross-terms from the internal metric, at the price of introducing $\eta$-dependence elsewhere and not only through the \SL(2) twist $A(\eta)$.

\subsection{Axionic flat directions and Wilson loops}\label{sec:wilson flat dirs}

It is natural to compare the $D=5$ Wilson loop interpretation of the cross-terms found in the internal metric of our solution with the uplift of some known flat directions associated to $D=4$ axion deformations of other $S$-fold solutions \cite{Guarino:2020gfe,Giambrone:2021zvp,Guarino:2021kyp,Guarino:2021hrc}.
The vacuum solutions of $[\SO(6)\times\SO(1,1)]\ltimes\bbR^{12}$ gauged supergravity have axionic flat directions  associated to \E7 generators analogous to $(t^{(\nu_a)})^T$, with 
\begin{equation}
\Sigma^{\rm axions}_{ABCD}  \propto \chi^{IJ} \epsilon_{IJ78\,ABCD}\,,
\end{equation}
where $\chi^{IJ}$ parametrise \su(4).
They generate $\bbR^{15}\subset\bbR^{27}$ and include as special cases the $\tilde\nu_a$ axions described above.
They uplift to the $\dd\eta$ components of the \SO(6) gauge connection in $d=5$ gauged maximal supergravity and are unaffected by the \SL(2) twist $A(\eta)$ in the uplift process.
The flat directions correspond to $\chi^{IJ}$ taking constant values along an abelian subalgebra of the \emph{residual} gauge symmetry. 
They have been shown in \cite{Giambrone:2021zvp,Guarino:2021kyp} to introduce a non-trivial fibering of $S^5$ over $S^1$ when uplifted to ten dimensions.

If we focus instead on their uplift to $D=5$ \SO(6) gauged maximal supergravity and follow the same reasoning as in the previous section, we notice that they give rise to Wilson loops along $S^1$, associated to an abelian subgroup of the {residual} gauge symmetries.
The fact that the associated symmetries are preserved, rather than broken, is the crucial difference compared to the vev \eqref{one form vev} found in our solution.
These Wilson loops are only invariant under gauge transformations globally defined on $S^1$ and a trivial Wilson loop means that the constant value of the associated vector field can be gauged away globally.
This time, no $\eta$ dependence will be introduced in other fields, since in this case the gauge symmetry being used is preserved.
This gives a straightforward interpretation of the periodicity of these axionic flat directions \cite{Giambrone:2021zvp,Guarino:2021kyp}, as the Wilson loops depend periodically on the axion vevs.

One may gauge away the vector vevs and associated Wilson loops even when they are non-trivial and the necessary transformation is multi-valued on $S^1$.
Since all non-vanishing matter fields are singlets under such gauge transformation, the end result is still a consistent field configuration, in contrast with the situation we described in the previous section.
The ten-dimensional interpretation of such a transformation is a multi-valued, $\eta$-dependent diffeomorphism on $S^5$ which removes the cross-terms between $S^5$ and $S^1$ in the local expression of the internal metric, but causes $S^5$ to be non-trivially fibered over $S^1$, where going around $S^1$ the deformed $S^5$ is twisted by a preserved isometry.
This matches the results in \cite{Giambrone:2021zvp,Guarino:2021kyp}.
{On the other hand, if we instead choose to compactify back from five to four dimensions, the gauge transformation used to remove these vector vevs in $D=5$ can be reinterpreted as introducing an \SO(6) twist in the KK ansatz, reproducing the interpretation in \cite{Guarino:2021hrc} that axion flat directions can be reabsorbed into the $D=4$ embedding tensor by adding extra Cremmer--Scherk--Schwarz couplings.}

This discussion can be summarised by stating that any gauged supergravity solution with vanishing vector field strengths and topology including an $S^1$ admits Wilson loop deformations along the circle, associated to the maximal torus of the residual gauge group, and that, if the solution can be reproduced from Kaluza--Klein reduction on $S^1$ (with or without duality twists), such deformations will manifest as axionic flat directions in the truncated model.

We can of course immediately apply this observation to our solution, by turning on a Wilson loop along $S^1$ for a generator of the residual \SO(3) gauge symmetry.
From the four-dimensional point of view, recall that the $D=4$ scalar mass spectrum \eqref{scalar masses} contains three massless scalars (apart from Goldstone bosons).
We identify these scalars with the subset of $\chi^{IJ}$ sitting in the adjoint of the residual gauge group (antisymmetrisation is understood):
\begin{equation}
\chi^{IJ} = \begin{pmatrix}
\chi^{\sf ij} & \\ & \chi^{\sf ij}
\end{pmatrix}\,,\qquad \mathsf i,\, \mathsf j = {\sf1,\,2,\,3}\,.
\end{equation}
Denoting $\chi^{\sf ij} t^{\rm axion}_{\sf ij}$ these \e7 generators, the solutions \eqref{N=1 solution 1}, \eqref{N=1 solution 2} are generalised to
\begin{equation}\label{sol with flatdir}
\cV_M{}^{\ul N} = \big( e^{\chi^{\sf ij} t^{\rm axion}_{\sf ij}} \cV_{\rm \SO(3)\,sol.}     \big)_M{}^{\ul N}
\end{equation}
with $\cV_{\rm \SO(3)\,sol.} $ the coset representative \eqref{N=1 coset param} evaluated at \eqref{N=1 solution 1} or \eqref{N=1 solution 2}.
Clearly, the \SO(3) gauge symmetry is broken down to \U(1) for generic values of $\chi^{\sf ij}$, while a quick computation shows that $\cN=1$ supersymmetry is preserved everywhere along the flat direction.

\section{Discussion}\label{sec:discussion}

In \cite{Arav:2020obl,Arav:2021tpk} a new numerical Janus solution of $D=5$, \SO(6) gauged maximal supergravity was found, which is periodic along the radial direction, corresponding to the $\eta$ direction here.
Being periodic allows one to compactify it on a circle, possibly up to an elliptic \SL(2,\bbZ) duality twist, giving rise to a solution of type IIB supergravity on $\mathrm{AdS}_4\times S^5\times S^1$.
The solution preserves $\cN=1$ supersymmetry and the same \SO(3) isometries of $S^5$ as the solution we presented here.
It is therefore natural to ask if our analytic solution is the same as the numerical one found in \cite{Arav:2020obl,Arav:2021tpk}.
We can see that this is not the case.
First of all, as pointed out already in \cite{Arav:2021tpk}, their solution cannot arise from a $D=4$ gauged supergravity, because such uplifts necessarily give rise to a constant warp factor in five dimensions, while the solutions studied there exhibit non-constant warp factors.\footnote{Another part of the argument given in \cite{Arav:2021tpk} is that solutions arising from $D=4$ gauged supergravity should exhibit a dilaton that depends linearly on the radial coordinate, rather than being periodic. This however only holds for uplifts of the $[\SO(6)\times\SO(1,1)]\ltimes\bbR^{12}$ gauging, which rely on a hyperbolic duality twist.}
The reason is rather simple: uplifting from four to five dimensions, all $\eta$ dependence is encoded in the \SL(2) duality twist $A(\eta)$, which is contained within $\E6$. 
But the warp factor is part of the five-dimensional Einstein frame metric, which is \E6 invariant.
We can indeed compute the $D=5$ metric from our solution and find%
\footnote{For instance, one can begin by taking \eqref{IIB metric} and applying a KK decomposition to identify the external metric of $D=5$ \E6 ExFT, which will depend on the five-sphere coordinates only through a power of the determinant of its round metric.
This factor is eliminated following the standard generalised Scherk--Schwarz ansatz on $S^5$. 
The overall normalisation is fixed so that the main AdS$_5$ vacuum of \SO(6) gauged maximal supergravity has radius $L^2$, reflecting the corresponding type IIB solution. Notice that we are using the same symbol as for the AdS$_4$ radius of our solution in section~\ref{sec:SO3 inv sol}. This should not cause any confusion.}
\begin{equation}
\dd s^2_{\rm 5d} \ =\ \frac{25\, L^2}{54} \left(\dd s^2_{\rm AdS_4} + \frac25 \dd\eta^2 \right)\,,
\end{equation}
with $\dd s^2_{\rm AdS_4}$ the ${\rm AdS_4}$ metric of unit radius.
Furthermore, this result is enough to evaluate the ratio of the effective five- and four-dimensional Newton's constants and compute then the free energy at large $N$.
We find
\begin{equation}
\cF_{S^3} \approx \left(\frac{2\pi}{n}+2\pi k\right)\,N^2\, \left(\frac{25}{54}\right)^{3/2}\!\sqrt{\frac25}\
\approx \left(\frac1n + k\right)\times 1.25178\, N^2\,,
\end{equation}
which, for $n=1$ and $k\ge0$, differs from the free energy of the periodic solution in \cite{Arav:2021tpk}.
Another difference between the two solutions is that our expression includes a non-trivial Wilson loop along $S^1$, associated with the gauge connection of a broken $\U(1)\subset\SO(6)$ gauge symmetry, as discussed in section~\ref{sec:broken sym wilson}.

We summarise a few basic observations on the CFT duals of the new solutions described in this paper.
In the globally geometric cases ($n=1$), we expect that the dual field theory must arise from the IR limit of a Janus-like configuration of $\cN=4$ SYM, where the complex coupling varies periodically along one direction that gets compactified to a circle.
A family of possible profiles for the complex coupling is given by \eqref{axiodil profile 1} for any fixed $\theta$ and $\omega$.
There may not be a relation between such configurations and interfaces of $\cN=4$ SYM, as there are neither well-defined asymptotic values for the complex coupling prior to compactification, and no duality twist is required afterwards.
When $n\neq1$, the resulting S-fold geometries will require an elliptic \SL(2,\bbZ) duality twist of $\cN=4$ SYM compactified on a circle. 
The axio-dilaton profiles in \eqref{axiodil profile 1}--\eqref{axiodil profile 3} again provide a family of complex coupling profiles along the compactified direction that one may use as a starting point to investigate these configurations.

There are other directions of investigation that open up from the results presented here.
Despite using a parametrisation of $S^5$ tailored to the properties of our solution, the ten dimensional expressions we have derived are quite complex, and it would be interesting to search for further simplifications.
We have briefly described the flat direction of our solution associated with the breaking of \SO(3) to \U(1) by a $D=5$ Wilson loop, noting it will correspond to a non-trivial fibering of $S^5$ over $S^1$ in analogy with \cite{Giambrone:2021zvp,Guarino:2021kyp}.
A more explicit study of this flat direction and its implications for a dual CFT is desirable.
Exceptional field theory techniques allow to study the spectrum of Kaluza--Klein excitations around solutions arising from generalised Scherk--Schwarz ans\"atze \cite{Malek:2019eaz,Malek:2020yue} and it would certainly be interesting to apply these techniques to our new solutions.
There may also simply be more vacuum solutions to be found in $\U(4)\ltimes\bbR^{12}$ gauged supergravity and other related models admitting uplifts to type II supergravities \cite{Inverso:2016eet}.
Finally, let us remark that although the solution presented here is completely analytic, its discovery came about through the numerical searches based exploiting modern auto-differentiation methods described in \cite{NUMPAPER}. This demonstrates the power of such numerical searches.  In fact, in \cite{NUMPAPER} another supersymmetric vacuum of the $\U(4)\ltimes\bbR^{12}$ model is found, along with several other non-supersymmetric ones. Their analytic expressions and possible uplifts should be investigated further.

\section*{Acknowledgements}
We would like to thank Nikolay Bobev and Fri\eth{}rik Gautason for discussions and correspondence.
G.I. thanks Adolfo Guarino for discussions.
This project has received funding from the European Union’s Horizon 2020 research and innovation programme under the Marie Skłodowska-Curie grant agreement No 842991.
DSB gratefully acknowledges the support by Pierre Andurand.

\appendix
\section{\texorpdfstring{\E7}{E₇₍₇₎} parametrisation}
\label{app:e7 param}

We use the \SL(8,\bbR) decomposition $\bf56\to\mathbf{28}+\mathbf{28}'$ of the \E7 fundamental, so that $V^M \to (V^{AB},\,V_{AB})$ leaving antisymmetrisation  understood.
Then, the \E7 generators are parametrised as follows
\begin{equation}\label{e7 gens sl8 basis}
t_M{}^N = \begin{pmatrix}
t_{AB}{}^{CD} & t_{AB}{}_{CD} \\ t^{AB}{}^{CD} & t^{AB}{}_{CD}
\end{pmatrix}
= \begin{pmatrix}
2\, \delta_{[A}{}^{[C} \Lambda_{B]}{}^{D]} &
 \Sigma_{ABCD}  \\
 \frac1{24} \epsilon^{ABCDEFGH}\Sigma_{EFGH}  &
-2 \, \delta_{[C}{}^{[A} \Lambda_{D]}{}^{B]} 
\end{pmatrix}\,,
\end{equation}
with $\Lambda_A{}^B$ the \SL(8,\bbR) generators and $\Sigma_{ABCD}$ fully antisymmetric.
The symplectic invariant reads
\begin{equation}
\Omega_{MN}=\begin{pmatrix}
 & \delta_{AB}{}^{CD} \\
  -\delta^{AB}{}_{CD}  &
\end{pmatrix}
\end{equation}

The generators associated to the $\cN=1$ truncation then are as follows, with self-explanatory notation.
The dilatons are associated to
\begin{align}
(\Lambda^{(\mathsf s)}  )_A{}^B &= \mathrm{diag}(+1,+1,+1,-1,-1,-1,-1,+1)\,, \\\nonumber
(\Lambda^{(\mathsf t_1)})_A{}^B &= \mathrm{diag}(+1,-1,-1,-1,+1,+1,-1,+1)\,, \\\nonumber
(\Lambda^{(\mathsf t_2)})_A{}^B &= \mathrm{diag}(-1,+1,-1,+1,-1,+1,-1,+1)\,, \\\nonumber
(\Lambda^{(\mathsf t_3)})_A{}^B &= \mathrm{diag}(-1,-1,+1,+1,+1,-1,-1,+1)\,, \\\nonumber
(\Lambda^{(\mathsf u_1)})_A{}^B &= \mathrm{diag}(+1,-1,-1,+1,-1,-1,+1,+1)\,, \\\nonumber
(\Lambda^{(\mathsf u_2)})_A{}^B &= \mathrm{diag}(-1,+1,-1,-1,+1,-1,+1,+1)\,, \\\nonumber
(\Lambda^{(\mathsf u_3)})_A{}^B &= \mathrm{diag}(-1,-1,+1,-1,-1,+1,+1,+1)\,, 
\end{align}
and we denote $(t^{(\mathsf x)})_M{}^N$ the embedding of these generators into the $\bf56$ according to \eqref{e7 gens sl8 basis}, with $\mathsf x$ running over the seven dilatons.
{To avoid confusion in the transition between single- and double-index notation, we state the normalisation $(t^{(\mathsf x)})_M{}^N(t^{(\mathsf x)})_M{}^N=24$ (no sum over $\mathsf x$).}
The axions are then associated to 
\begin{equation}\label{sigmas of axions}
\begin{array}{rlrlrl}
\Sigma^{(\sigma)}_{ABCD}  &\!\!= + 24 \,\, \delta^1_A{}^2_B{}^3_C{}^8_D\,,  \qquad     &
\Sigma^{(\tau_1)}_{ABCD}  &\!\!= - 24 \,\, \delta^1_A{}^5_B{}^6_C{}^8_D\,,    &
\Sigma^{(\nu_1)}_{ABCD}   &\!\!= - 24 \,\, \delta^1_A{}^4_B{}^7_C{}^8_D\,,    
\\[1.5ex]
                          &                                               & 
\Sigma^{(\tau_2)}_{ABCD}  &\!\!= + 24 \,\, \delta^2_A{}^4_B{}^6_C{}^8_D\,, \qquad   &
\Sigma^{(\nu_2)}_{ABCD}   &\!\!= - 24 \,\, \delta^2_A{}^5_B{}^7_C{}^8_D\,,    
\\[1.5ex]
                          &                                               &
\Sigma^{(\tau_3)}_{ABCD}  &\!\!= - 24 \,\, \delta^3_A{}^4_B{}^5_C{}^8_D\,,    &
\Sigma^{(\nu_3)}_{ABCD}   &\!\!= - 24 \,\, \delta^3_A{}^6_B{}^7_C{}^8_D\,.
\end{array}
\end{equation}
We denote $(t^{(\chi)})_M{}^N$ the embedding of these generators into the $\bf56$ according to \eqref{e7 gens sl8 basis}, with $\chi$ running over the seven dilatons.
The normalisation is \mbox{$(t^{(\chi)})_M{}^N(t^{(\chi)})_M{}^N=12$}.
The coset representative for the $(\bbZ_2)^3$ invariant $\cN=1$ truncation is then given by the following expression
\begin{equation}\label{N=1 coset param}
\cV_M{}^{\underline N} =  
e^{-\sigma \, t^{(\sigma)}} \,{\sf s}^{\frac12 t^{(\mathsf s)}}  \ 
\prod_{a=1,2,3} e^{-\tau_a \, t^{(\tau_a)}} \,({\sf t}_a)^{\frac12 t^{(\mathsf t_a)}}\ 
\prod_{a=1,2,3}e^{-\nu_a \, t^{(\nu_a)}} \, ({\sf u}_a)^{\frac12 t^{(\mathsf u_a)}} \ .
\end{equation}

\providecommand{\href}[2]{#2}\begingroup\raggedright
\endgroup

\end{document}